\documentclass[3p]{elsarticle}

\makeatletter
\def\ps@pprintTitle{%
 \let\@oddhead\@empty
 \let\@evenhead\@empty
 \def\@oddfoot{\centerline{\thepage}}%
 \let\@evenfoot\@oddfoot}
\makeatother

\usepackage{moreverb,url}
\usepackage{graphics}
\usepackage{graphicx}
\usepackage{epsfig}
\usepackage{multirow}
\usepackage{amssymb}
\usepackage{amsthm}
\usepackage[colorlinks,bookmarksopen,bookmarksnumbered,citecolor=red,urlcolor=red]{hyperref}
\usepackage{subcaption}
\usepackage{natbib}
\usepackage{amsmath}
\usepackage{booktabs}
\usepackage{longtable}
\usepackage{xfrac}
\usepackage[dvipsnames]{xcolor,colortbl} 

\usepackage{float} 
\usepackage[normalem]{ulem} 
\usepackage{natbib}
\usepackage{setspace}
\usepackage{wrapfig}

\usepackage{chngcntr} 

\usepackage{bm} 
\usepackage{units} 

\usepackage{tikz} 
\newcommand*\circled[1]{\tikz[baseline=(char.base)]{
            \node[shape=circle,draw,inner sep=2pt] (char) {#1};}}

\usepackage{wasysym} 

\begin{document}

\begin{frontmatter}

\title{From Playground Swings to Sway Control of Cranes: An Active Pendulum Experiment}
\author[mysecondaryaddress]{Adrian Stein}
\author[mysecondaryaddress]{Tarik Parcic}
\author[mysecondaryaddress]{Tarunraj Singh}
\cortext[mycorrespondingauthor]{Corresponding author}
\ead{tsingh@buffalo.edu}
\address[mysecondaryaddress]{Department of Mechanical and Aerospace Engineering, University at Buffalo (SUNY),\\ Buffalo, NY 14260-4400, USA}

\begin{abstract}
Dynamics is a core discipline in Mechanical and Aerospace Engineering programs and with the ubiquitous nature of control in modern day applications, the field of mechatronics has gained popularity. Mechatronics  refers to the field of engineering which integrates the engineering disciplines of mechanical, control, electronics and computing. To create a testbed to illustrate a tabletop mechatronics system, the paper details the design, and fabrication of an active pendulum whose length can be changed in real-time using solenoids. This permits illustrating two concepts: (1) damping of pendulum oscillations which emulates the sway of a crane and (2) amplification of the oscillations which emulates the pumping of a playground swing. The paper describes the steps prior to experimental validation which include: modeling, system identification, signal processing, and controller implementation. Numerical simulations are used to prototype the controller and eventually to compare the simulation results to the experimental ones. The results of all the experiments illustrate a close match between the simulated and experimental results. To permit reproduction of the experiment, the design details and code to implement the controllers are posted in a public repository.
\end{abstract}


\end{frontmatter}

\section{Introduction}
\label{sec:introduction}
\begin{figure}
	\centering
	\includegraphics[width=0.6\textwidth]{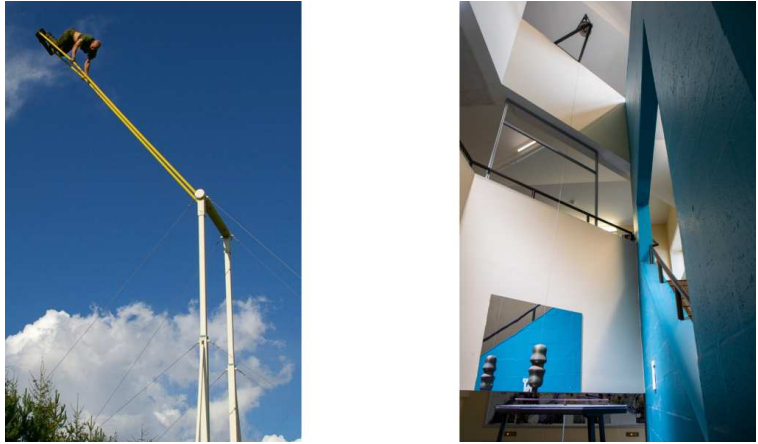}
    \caption{Kiiking, Foucault Pendulum at UB: Exponents of Pendulums (distributed under a \href{https://creativecommons.org/licenses/by-sa/3.0/}{CC BY-SA 3.0 license.})}
	\label{fig:Kiiking}
\end{figure}

Kiiking (Figure~\ref{fig:Kiiking}) is a popular sport in Estonia (also called extreme swinging), where the cables/chains of a playground swing are replaced with rigid rods to permit the swing to rotate full circle. The swing is pumped up by changing the center of mass at specific time instants based on feedback of the position and velocity extremes. This form of pumping up a swing has been extensively studied with Wirkus et al.~\cite{WirkusRandR.&RuinaA..1998} modeling a playground swing as a pendulum and present elegant analysis of two approaches to pump up a swing. They first present a physics based understanding of the Kiiking strategy followed by the analysis of the intuitive strategy that most of us use to pump up the swing from a sitting position. It is particularly noteworthy that the pumping strategy adopted by children on playground swings is the time-optimal solution to maximizing the amplitude of oscillation~\cite{Piccoli.2005}. Further, Glendinning~\cite{glendinning2018shaking} noted the strategy of a child pumping up a swing is analogous to the web-shaking of {\it Agriope aurantia}, an orb-web spider, which could be a defense mechanism by making it harder for a predator to strike due to the rapid change in the location of the spider. This serves as another piece of evidence of the maxim that nature optimizes for specific goals.
The pendulum historically has been used for keeping time and for keeping a beat as a metronome.  It is also a widely studied physical example to teach students dynamics and controls. 
Study of pendulums has a rich history which includes Galileo Galilei's work on pendulums which resulted in observations leading to the development of the pendulum clock by Huygens~\cite{stillman2001galileo} in 1657. Galileo's study of a pendulum resulted in the conclusion that the period of oscillation is independent of the displacement (isochronism), which in reality is only approximately the same. Galileo as a student had conjectured that Aristotle’s conclusion that a heavier body would fall faster than a lighter one, was incorrect. He also illustrated that the mass of the pendulum did not change the period of oscillation which helped support his assertion that bodies of different masses take the same time to fall a given distance. Towards the end of the Apollo 15 moon walk, Commander David Scott of the Apollo 15 mission demonstrated Galelio's assertion when he dropped a hammer and a feather, and the objects struck the surface at the same time~\cite{NASA15}.

From its profound impact for maritime navigation where a pendulum clock was used to solve ``the longitude problem''~\cite{Sobel.2007}, to its accessible illustration of the Earth's rotation via the Foucault's pendulum~\cite{baker2008pendulum}, to its use to model the Segway Robot as an inverted pendulum~\cite{castro2012modeling}, it serves as a benchmark problem at various levels of education. It is interesting to note that the John Harrison's clock~\cite{Sobel.2007} demonstrated that it could be used to determine longitude and the Foucault's pendulum's (Figure~\ref{fig:Kiiking}) precession rate can be used to determine the latitude.
 \begin{figure}
	\centering
	\includegraphics[width=.7\textwidth]{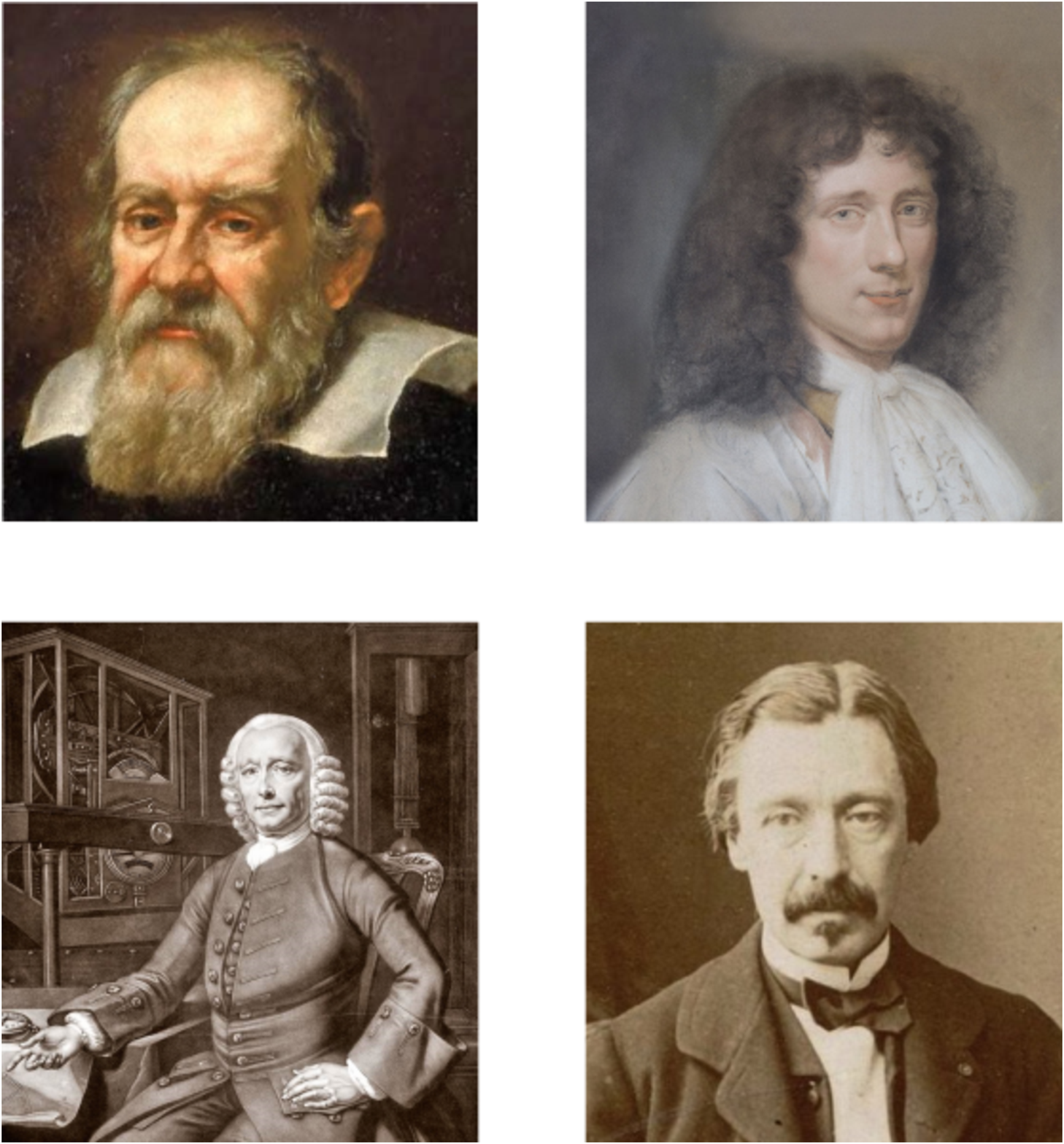}
	\caption{From left-to-right and top-to-bottom: Galileo Galilei, Christiaan Huygens, John Harrison, L\'{e}on Foucault: Exponents of Pendulums (distributed under a \href{https://creativecommons.org/licenses/by-sa/3.0/}{CC BY-SA 3.0 license.}).}
	\label{fig:figure3}
\end{figure}
Strogatz in his book {\it SYNC: The Emerging Science of Spontaneous Order}~\cite{Strogatz.2003} presents numerous examples of synchrony in nature and submits that at the heart of comprehending synchrony is the study of ``coupled oscillators''. Using examples such as groups of fireflies, pacemaker cells, wobble of the Millennium Bridge~\cite{eckhardt2007modeling} etc., he motivates the use of coupled oscillators to understand the mesmerizing phenomena of synchrony. The idea that two oscillators can synchronize was first demonstrated by Huygens in 1664~\cite{willms2017huygens} when he noted that two pendulums of equal length hung from a beam fell in-phase in about 30 minutes, i.e., became synchronous. Again, the commonplace pendulum creating the science of coupled oscillators.

The rich dynamics of the simple pendulum has resulted in numerous papers over the past half century. Curry~\cite{Curry.1976} characterized the dynamics of pumping up a swing as a parametric amplifier and noted that the rate of growth of energy was independent of the mass of the child on the playground swing. Stilling and Szyszkowski~\cite{Stilling.2002} estimated the equivalent damping of a pendulum where the length can be changed dynamically in terms of the rate of change of the length. For small displacement of the pendulum, they approximated the damping ratio to be a function of the magnitude of perturbations of the length of the pendulum about a nominal position, confirming the assertion of Curry~\cite{Curry.1976}, that the rate of change of the amplitude of oscillation is independent of the mass of the pendulum. Vyhl\'idal et al.~\cite{Vyhlidal.2017} demonstrated via an experiment that the variable length pendulum can generate attenuation of the oscillation of a pendulum with a damping ratio $\leq 0.1$.

Given the ubiquity of pendulums from the Black Forest novelty Kuckucksuhr (Cuckoo Clock) to cranes, access to a tabletop crane experiment which could be used to demonstrate pumping up a swing or actively damping the sway of cranes would serve as a wonderful testbed to motivate K-12 students and undergraduate students interested in dynamics and control. Tea and Falk~\cite{Tea.1968} in  their 1968 paper describing the {\it Pumping on a Swing}, state that many undergraduate physics book ``miss a good pedagogic opportunity by not including this topic'', a sentiment that we endorse. They develop a model based on conservation of angular momentum to determine the change in the magnitude of displacement of the swing. Besides the aforementioned motivations, Belendez et al.~\cite{belendez2007exact} derive closed form expressions for the motion of a pendulum using the exact nonlinear equations using Jacobi elliptic functions. They also derive a closed form equation which illustrates how the frequency of oscillation of the pendulum changes with the maximum displacement, a nonlinear phenomena which is not generally discussed in undergraduate classes. The estimation of the frequency of oscillation as a function of the initial displacement will provide a simple yet powerful experimental illustration of nonlinearities and how linear models can only function around some operating point and large deviations from the operating points lead to the deterioration of the fidelity of the model.

The objective of this paper is to provide a detailed exposition of the design and fabrication of a solenoid driven pendulum which can be used to illustrate both the concept of pumping up a swing, and damping the sway oscillations of a pendulum, which serve as a simple illustrations of positive and negative feedback.

Teh et al.~\cite{Teh.2015} developed an experimental rig to illustrate the dynamics of a parameterically excited pendulum where the pivot of the pendulum is subject to an oscillatory motion. A solenoid provides a force to counter gravity and a spring ensures that the gravity provides the complementary force without free fall. A period-1 and period-2 oscillation is demonstrated based on different forcing frequencies.
In contrast to the focus of the parameterically excited problem, the design proposed in this paper is to illustrate pumping up a swing and damping pendular motion using solenoids. The intent is not to result in a parameterically excited pendulum, rather to permit changing its period of oscillation dynamically.
\subsection{Experimental setup}
\begin{figure}
	\centering
	\includegraphics[width=\textwidth]{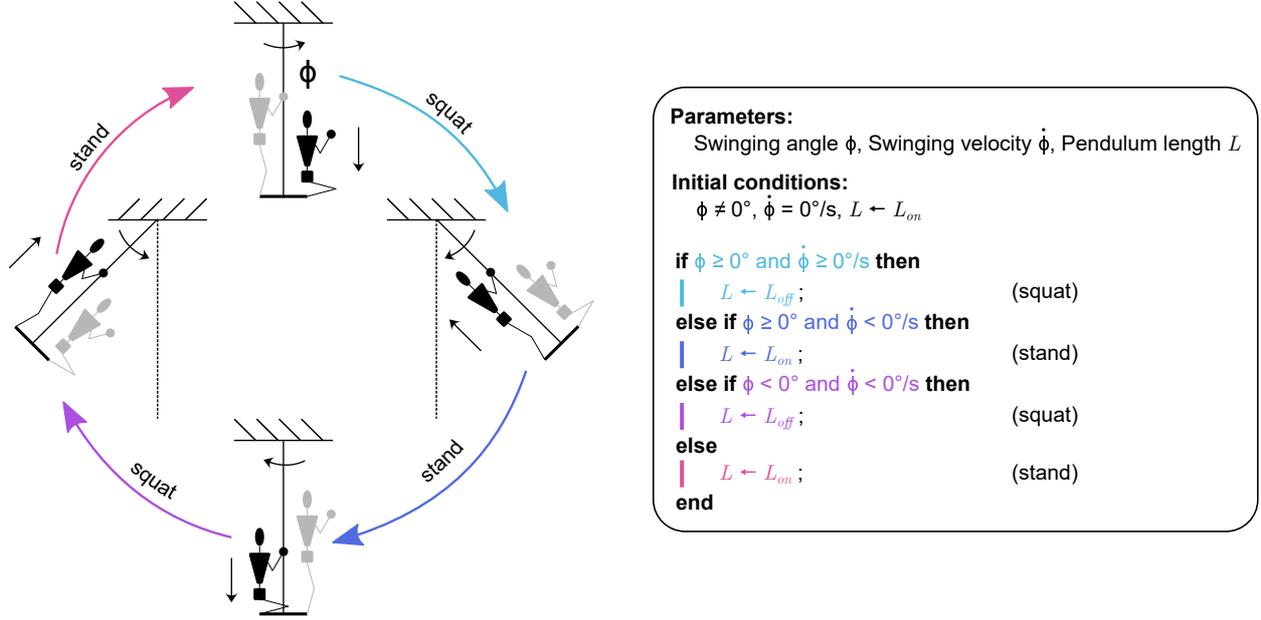}
	\caption{Attenuation of a pendulum oscillation illustrated for a person on a swing. Angle $\Phi$ denotes the rotation of the pendulum, where counter-clockwise is the positive rotation ($\Phi=0$ $^\circ$ at the vertical equilibrium). Change from standing to squatting at $\Phi=0$ $^\circ$ and from squatting to standing at $\dot{\Phi}=0$ $^\circ$/s.}
	\label{fig:01}
\end{figure}

\begin{figure}
	\centering
	\includegraphics[width=0.95\textwidth]{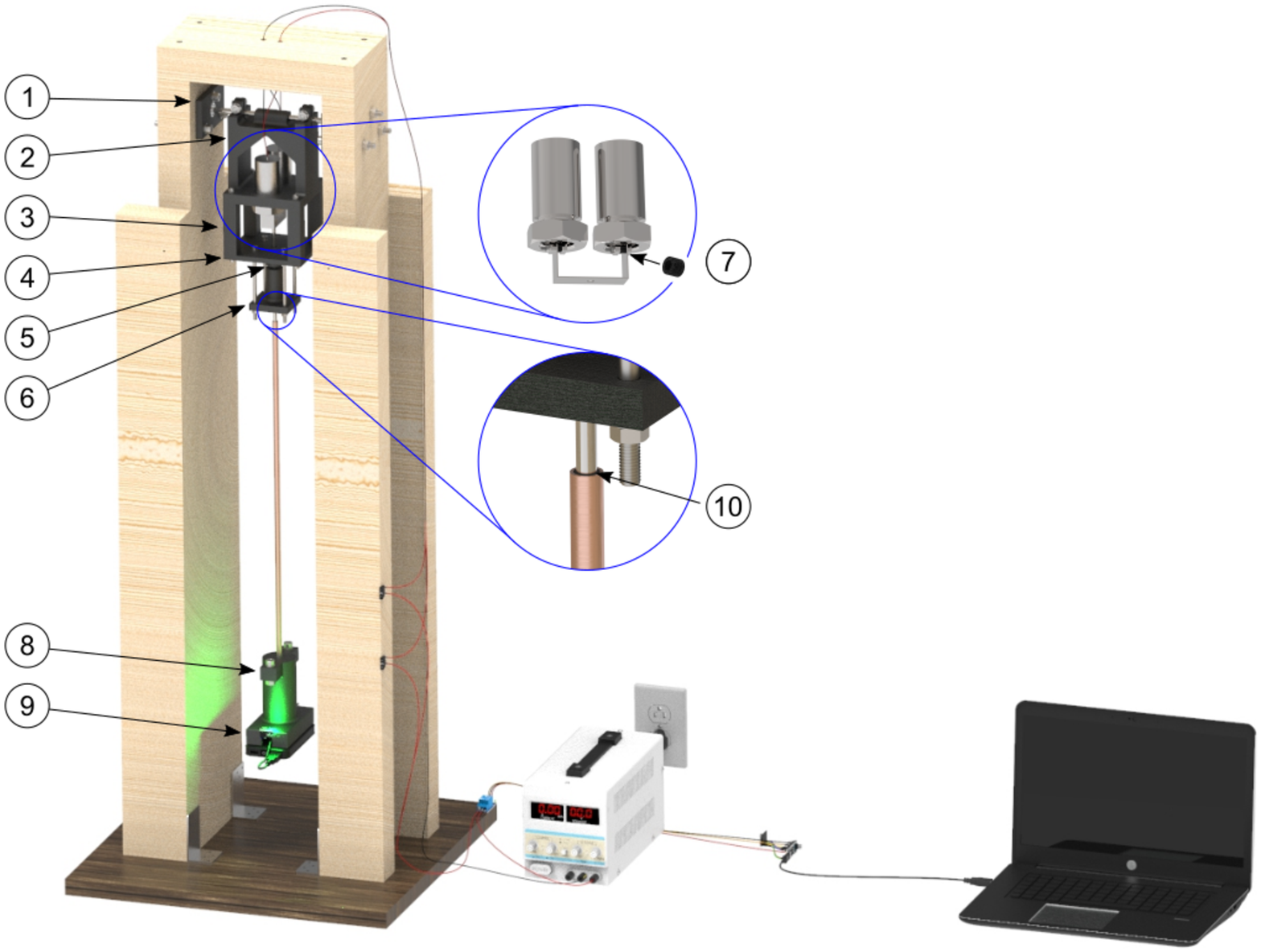}
	\caption{General setup of the pendulum with highlighting all the $3$D printed parts}
	\label{fig:04N}
\end{figure}

A one degree of freedom pendulum to emulate a playground swing is the focus of this work. To illustrate the act of pumping up a swing, an additional degree of freedom is included which permits changing the length of the pendulum via solenoids. The same setup can also be used to illustrate the active damping of sway motion of a crane.
Figure~\ref{fig:01} illustrates the active damping strategy for the attenuation of the sway of a pendulum, which shows that a person on a swing has to squat at the vertical equilibrium and then stand up when the displacement is maximal. The person maintains their position between the vertical and the extreme displacement. 
The energy dissipates only in the vertical position when the angle \mbox{$\Phi=0$ $^\circ$} and the person squats. Assume a surrogate model for a person on a swing is a pendulum where the lengths can be varied by switching solenoids on and off. At \mbox{$\Phi=0$ $^\circ$} 
the angular velocity \mbox{$\dot{\Phi}$ $^\circ$/s}
changes as follows \cite{WirkusRandR.&RuinaA..1998}:
\begin{align}
\dot\Phi^+ = \left(\frac{L_{on}}{L_{off}}\right)^2\dot\Phi^- \label{eq:phi_dot_plus}
\end{align}
where $L_{on}$ is the length of the pendulum when the solenoids are turned on and $L_{off}$ when they are turned off ($L_{off}$ $>$ $L_{on}$), while $\dot\Phi^+$ describes the angular velocity after the pendulum transition through the vertical and $\dot\Phi^-$ is the angular velocity prior to the transition. Note that turning the solenoids on reduces the length of the pendulum and gravity and centrifugal forces passively increase the length of the pendulum. The precise timing of turning the solenoids on and off will be described in the following sections. Figure~\ref{fig:04N} gives an overview of the active pendulum setup. The main parts are the pendulum with the solenoids, power module, sensors and the portable computer. Various parts of the pendulum setup were 3D printed and are listed in Table~\ref{table:3D_printedN}.
\begin{table}
\caption{$3$D printed parts for the experiment.}
\centering 
\small
\begin{tabular}{l c c c c c c} 
\toprule 
Part number & Part name & Quantity\\
\midrule
Bearing bracket & \circled{$1$} & $2$\\
Housing solenoid  & \circled{$2$} & $1$\\
Adapter spacer & \circled{$3$} & $4$\\
Mid base & \circled{$4$} & $1$\\
Spacer Bearing & \circled{$5$} & $1$\\
Lower base & \circled{$6$} & $1$\\
Spacer solenoid  & \circled{$7$} & $2$\\
Payload chassis cover & \circled{$8$} & $1$\\
Payload chassis & \circled{$9$} & $1$\\
Copper tube spacer & \circled{$10$} & $2$\\
\bottomrule
\label{table:3D_printedN}
\end{tabular}
\end{table}
A $3$D-CAD model was developed to permit breakdown of the components and provide a detailed illustration for the fabrication of the system. 
The solenoids change the length of the pendulum directly and are constantly swinging with the pendulum. Since the solenoid works instantaneously, it is perfectly suited for the fast displacement of the center of mass. When the solenoids are turned off, the pendulum is extended and contracted when the solenoids are turned on. 
The solenoids are integrated into a $3$D printed body and connected to the pendulum via a U-beam and are powered externally. The solenoid we used are model ``F0494A" from digikey. They have a pull length of $31.75$ mm and can be operated with a maximum of $10$ A. To increase the pulling power, two solenoids are installed in parallel so that their forces add up and permit lifting a greater mass. Note that the mass does not impact the frequency of oscillation of the pendulum. Since the tip mass includes a $5$ V power bank, gyro and wireless communication hardware, two solenoids were deemed necessary. The tip mass is connected to the solenoids via a metal rod which slides in a linear bearing attached to the structure housing the solenoids. The active pendulum setup is housed in a wooden frame attached to a base plate making it portable.
\section{System identification and signal processing}
\label{sec:system_identification}
This section deals with the collection of gyro measurement data and identifying the parameters of the system model, a process referred to as system identification. Since the gyro data can be contaminated by noise and high frequency structural vibration, it is low pass filtered.
\subsection{Low-pass filter}
\label{subsec:low_pass_filter}
\begin{figure}
	\centering
	\includegraphics[width=0.90\textwidth]{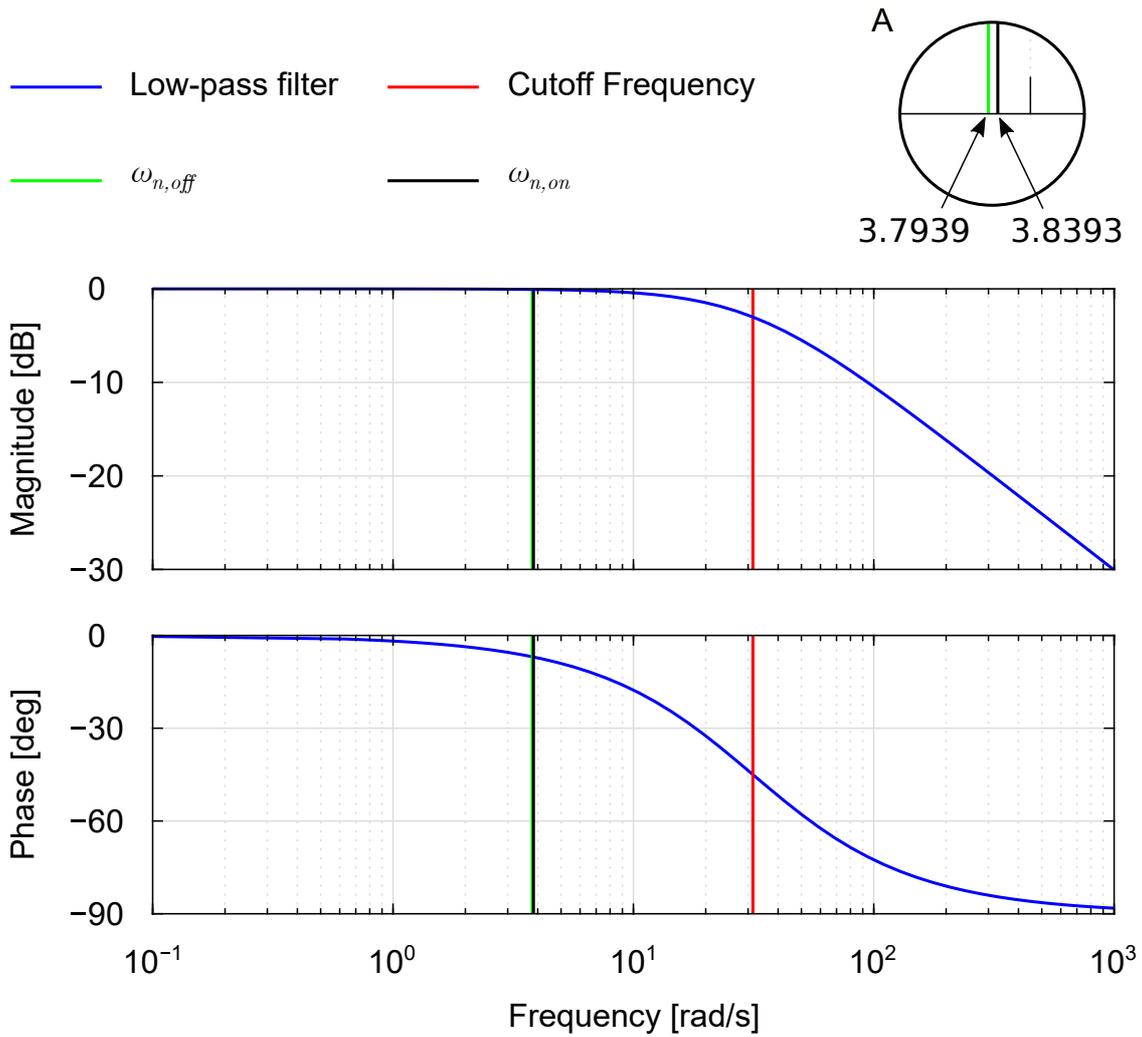}
	\caption{Low-pass filter with cutoff frequency at $31.4159$ rad/s ($5$ Hz). The green and the black line represent the natural frequencies of the pendulum for deactivated and activated solenoids respectively. Inset A provides a zoomed view on the cutoff and natural frequency.}
	\label{fig:19}
\end{figure}

Since the solenoid closes violently, it can excite unmodelled dynamics such as the structural modes of the pendulum rod. To acquire only the rotary motion of the pendulum without the high frequency components of the signal, a low-pass filter is designed and implemented. Figure~\ref{fig:19} illustrates the location of the natural frequencies (via the green and black lines) of the pendulum with the solenoid turned off and on. The cutoff frequency of the low-pass filter is represented by the red line.
The cutoff frequency is at $5$ Hz and is a trade off between attenuating higher frequencies and creating a phase shift as small as possible in the pass band.
\subsection{Calculation of natural frequency and damping ratio}
\begin{figure}
	\centering
	\includegraphics[width=0.75\textwidth]{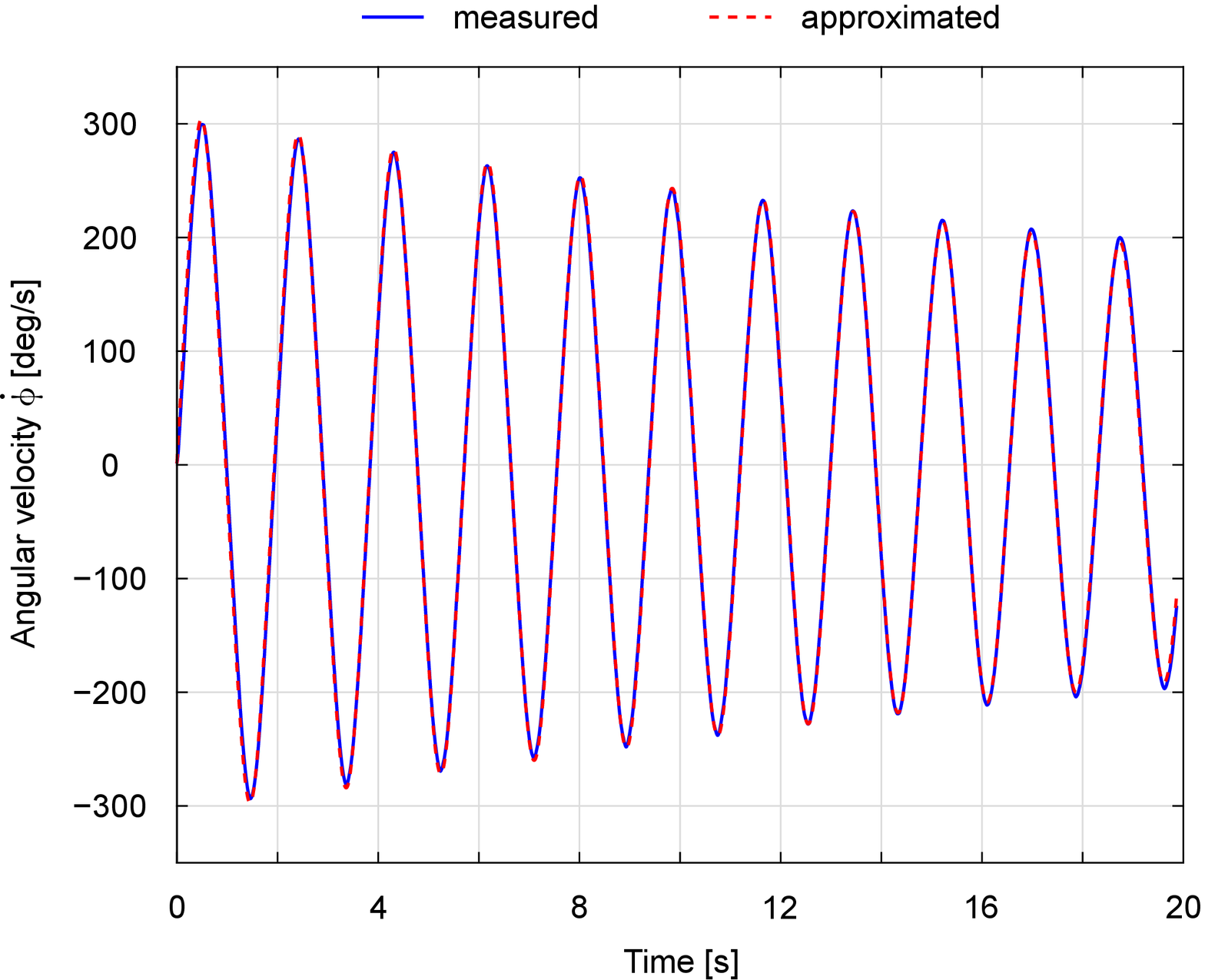}
	\caption{Measured $\dot{\Phi}$ (solid blue) from an experiment with permanently deactivated solenoids. Curved fitted $\dot{\Phi}$ (dashed red) from an equation of motion to determine the scaled damping constant $c$ and the length of the pendulum $L_{off}$ when the solenoids are turned off.}
	\label{fig:18}
\end{figure}

In contrast to \cite{Stilling.2002}, we assume that our pendulum dynamics are damped. The following model describes the equation of motion of the pendulum:
\begin{align}
    \ddot{\Phi} + c \dot{\Phi} + \omega_n^2\Phi & = 0\\
    c = 2\zeta\omega_n, \: \: \: \omega_n^2 = \frac{g}{L_{off}},
\end{align}
where $c$ is the scaled damping constant and $L_{off}$ is the length of the pendulum when the solenoid is deactivated. We displace the pendulum from its equilibrium position and leave the solenoids deactivated and let the pendulum swing freely. With the knowledge that $g = 9.81$ m/s\textsuperscript{2} and the initial conditions $\Phi(0)=-90^\circ$ (holding the pendulum in the beginning) and $\dot{\Phi}(0)$ provided by the gyro, the model parameters $c$ and $L_{off}$ are estimated by posing a least squares optimization problem. MATLAB's function {\it ode45} and {\it fmincon} functions are used to solve the least-squares problem from which we  determine the values for $c$ and $L_{off}$. With a Vernier caliper we measure the stroke of the solenoids and calculate $L_{on}$ from $L_{off}$. Figure~\ref{fig:18} shows the remarkable curve fit for our experimental results which validates the estimated model. 
In our case, the measured stroke of the solenoid is $16$ mm. It should be noted that the stroke can vary, depending on the manufacturing tolerances of the solenoids and the precision of the $3$D printed parts. After curve-fitting the measured data with our model we found that $\zeta = 0.0067$, $L_{off} = 0.6815$ m and $L_{on} = 0.6655$ m.
We can calculate the natural frequencies for the deactivated and activated case with:
\begin{align}
\omega_{n,(.)} = \sqrt{\frac{g}{L_{(.)}}} \label{eq:omega_n}
\end{align}
By using Eq.~\eqref{eq:omega_n} $\omega_{n,off} = 3.7939$ rad/s and $\omega_{n,on} = 3.8393$ rad/s. 
\section{Experimental validation}
\label{sec:experimental_results}
Experiments were conducted to illustrate the active damping performance of the length changing pendulum strategy and its dual which is pumping up the pendulum (swing). Simulations results are also included to illustrate how well the experimental results validate the simulated results.
\subsection{Active sway attenuation of the pendulum}
\begin{figure}
	\centering
	\includegraphics[width=0.75\textwidth]{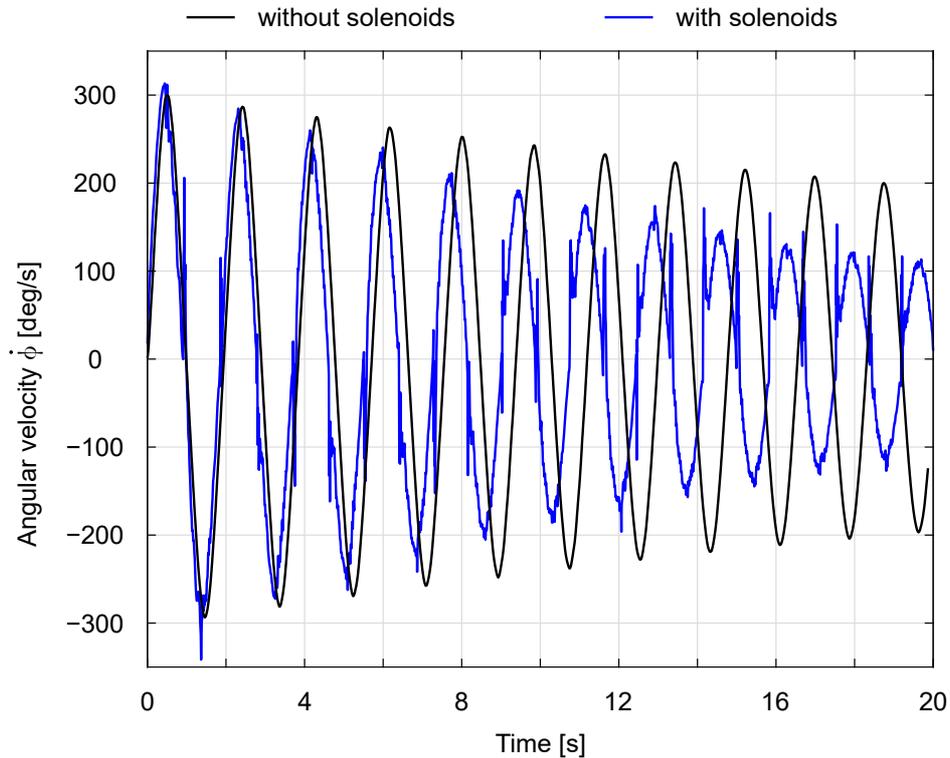}
	\caption{Angular velocity $\dot{\Phi}$ over time of a scenario with freely swinging pendulum (black) versus a scenario with the solenoids activated (blue) for attenuation.}
	\label{fig:20}
\end{figure}

This experiment was conducted to illustrate the damped response of an uncontrolled pendulum and compares it to the attenuation of the oscillations with an active control strategy. For both the experiments, the 
pendulum was displaced to \mbox{$\Phi(t=0s)=-90$ $^\circ$}. Figure~\ref{fig:20} illustrates the time response of two scenarios using the angular velocity of the pendulum. From the time response of the angular velocity $\dot{\Phi}$ of the pendulum, it can be noted that over 20 seconds, the amplitude of the actively controlled pendulum is about half of the uncontrolled response.
It can also be noted that there is an increasing phase shift in the response which is attributable to the time varying natural frequency of the pendulum.
\subsection{Simulated vs experimental results}
\begin{figure}
	\centering
	\includegraphics[width=0.75\textwidth]{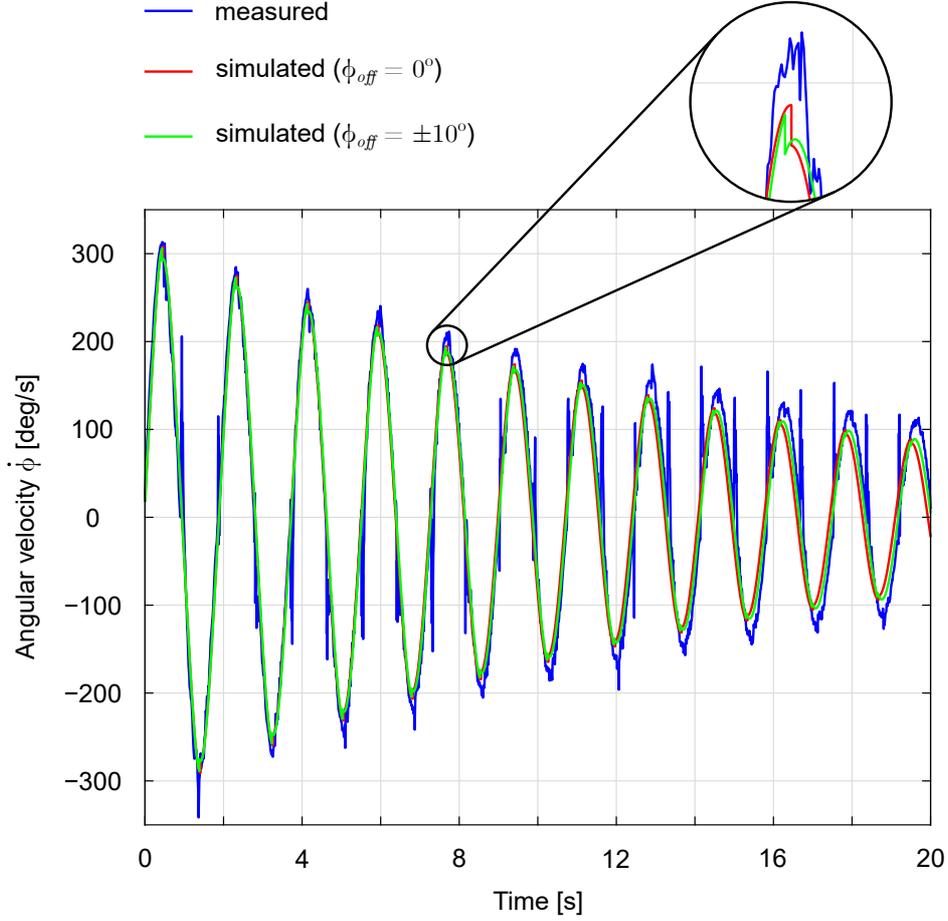}
	\caption{Angular velocity $\dot{\Phi}$ over time from the experiment (blue) versus results from simulations (red \& green). The red graph shows the simulation results when the solenoids are deactivated at $\Phi=0$ $^\circ$ while in the green graph they are deactivated at $\Phi=\pm 10$ $^\circ$ too early. The initial condition for the angle is $\Phi(0)=-90^\circ$.}
	\label{fig:21}
\end{figure}

To numerically mimic the change in the rod length of the pendulum, MATLAB's ode-event function was used. To precisely simulate the change in the pendulum length as a function of displacement and velocity of the pendulum, the event function was used to detect when $\Phi=0$~$^\circ$ or $\dot{\Phi}=0$~$^\circ$/s. When an event is triggered, the current values of $\Phi$ and $\dot{\Phi}$ are used as initial conditions for the subsequent numerical integration of the equations of motion. Figure~\ref{fig:21} illustrates that the experimental response $\dot{\Phi}$ (blue) is very close to the simulated results (red). When the solenoids are energized at $\dot{\Phi}=0$~$^\circ$/s some potential energy is gained but not as much as the loss of energy when $\Phi=0$~$^\circ$. Experimentally determining the exact time when $\Phi = 0$ $^\circ$ by estimating when $\ddot{\Phi} = 0$ $^\circ$, is not a feasible approach. This is attributed to the fact that the derivative of the rate information is not smooth because of the noise in the sensed data. The time of zero crossing of $\Phi$ was estimated based on the period of oscillation of the pendulum which is amplitude dependent. Furthermore, the transition time to actuate or release the solenoid is small, but not negligible which will impact the estimate of the period of oscillation of the pendulum. One could ask, how much do errors in timing of release of the solenoids (at $\Phi = 0$) impact $\dot{\Phi}$? As an exercise, we chose that the solenoids get deactivated $10$~$^\circ$ earlier than the ideal time, when transitioning forward or backward. This is to account for the inherent delays in numerical processing and releasing the solenoid. Figure~\ref{fig:21} illustrates that the impact is very small for such an uncertainty. The authors are aware of sophisticated techniques to estimate $\dot{\Phi}$ or $\Phi$, such as Kalman filtering. Implementing such observers are reserved for future research.
\subsection{Pumping up a pendulum}
\label{subsec:pushing_pendulum}
\begin{figure}
	\centering
	\includegraphics[width=0.75\textwidth]{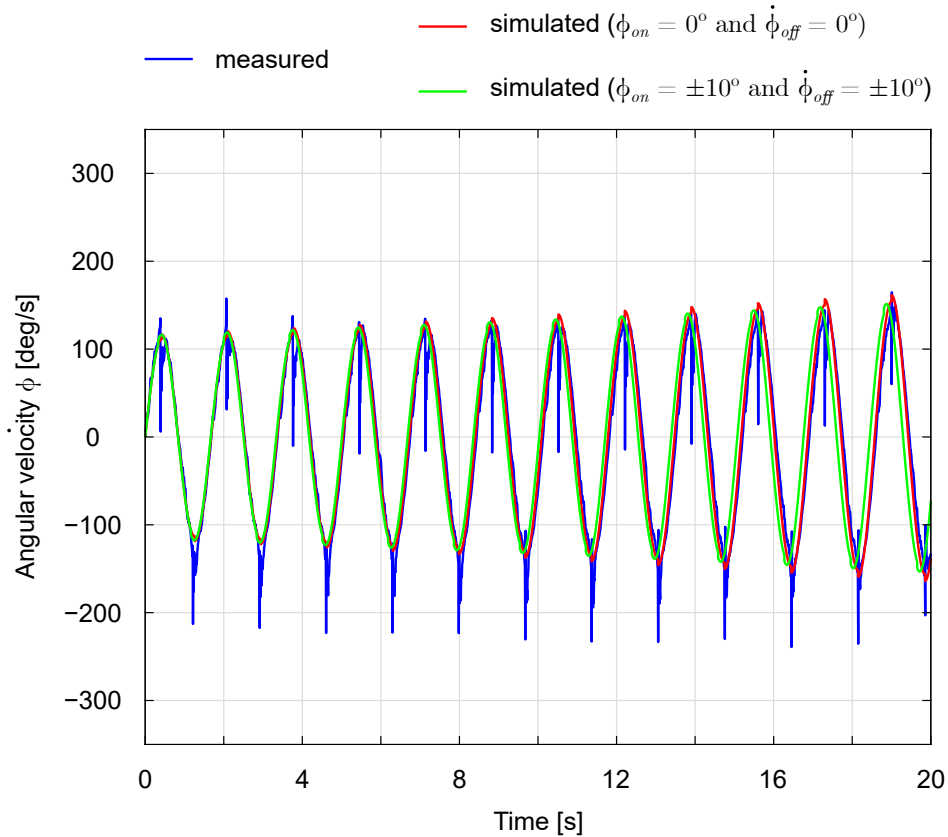}
	\caption{Angular velocity $\dot{\Phi}$ over time from the experiment (blue) versus results from simulations (red \& green). The red graph shows the simulation results when the solenoids are activated at $\Phi=0$ $^\circ$ and deactivated at $\dot{\Phi}=\pm 0$ $^\circ$/s while in the green graph they are activated $\Phi=\pm 10$ $^\circ$ too early and $\dot{\Phi}=\pm 10$ $^\circ$/s too late. The initial condition for the angle is $\Phi(0)=-30^\circ$.}
	\label{fig:22}
\end{figure}

By switching the strategy illustrated in Figure~\ref{fig:01} where the pendulum length is reduced when $\Phi = 0$~$^\circ$ and increased when $\dot{\Phi} = 0$~$^\circ$/s, one can emulate the strategy for pumping up a playground swing. It should be noted that the centrifugal force reaches its maximum when the velocity peaks. This for the active damping case adds to the force of gravity which is collinear with the pendulum to extend the pendulum passively when $\Phi = 0$~$^\circ$. For the pumping up the pendulum scenario, the solenoid force contracts the pendulum at $\Phi = 0$~$^\circ$ and has to fight the gravitational and centrifugal force, while at $\dot{\Phi} = 0$~$^\circ$/s, the centrifugal force is $0$ and the gravitational force along the rod's axis is just $\cos\left(\Phi\right)mg$ which progressively reduces the force to extend the pendulum passively as the displacement increases. An extreme case is when $\Phi=90$~$^\circ$, the gravitational force is orthogonal to the pendulum and concurrently the centrifugal force is zero resulting is no extension of the pendulum. This is due to the limitations of the current experimental setup. If opposing solenoids were used to retract and extend the pendulum, this limitation of the current setup could be eliminated. With the aforementioned constraints in mind, the pumping strategy is validated for small initial displacement of the pendulum. Figure~\ref{fig:22} illustrates the experimental and simulated results for small displacements in $\Phi$. For the simulation, the red graph assumes that the solenoid gets contracted at $\Phi = 0$~$^\circ$ and released at $\dot{\Phi} = 0$~$^\circ$/s. The green graph assumes that the solenoid gets contracted $\Phi = 10$~$^\circ$ too early and released $\dot{\Phi} = 10$~$^\circ$/s too late. We decided to choose a premature release to account for the limitations of the hardware. It can be seen that the green curve fits the measured data almost exactly, which confirms that the solenoids don't get activated at $\Phi = 0$~$^\circ$ and a delayed expansion of the pendulum is taking place.
\section{Conclusion}
The scope of this paper is the design and assembly of a tabletop experiment of an active pendulum to illustrate the concept of pumping up a playground swing and damping the sway motion of a crane. Since the pendulum serves as a surrogate model for numerous applications including cranes, slosh of fluid in containers, playground swing, and the inverted pendulum for devices such as a unicycle and self-balancing mobility platform such as the Segway, access to a relatively inexpensive design can serve to make it an accessible experiment for various engineering and physics courses. The core idea of the design presented in the paper is to displace the pendulum's center of mass at prescribed time instants, motivated by the well-known observation of children pumping up a swing on a playground, which has proven to be the time-optimal solution for maximizing the amplitude growth of the oscillations. In addition to pumping up a swing, our paper also addresses the question: How can a pendulum's motion be attenuated as fast as possible when using solenoids to change its length? The motivation of using solenoids as the actuator are: 1. The force is applied almost instantaneously, and 2. The switching control is relatively easy to implement. There are a couple of constraints which are noted:
\begin{itemize}
    \item The attenuation rate of the sway depends on the stroke length of the solenoids,
    \item The efficiency of damping the movement depends on the timely actuation of the solenoid,
    \item The mechanical impacts of the solenoid require filtering of the gyroscope's signal since it can excite structural modes.
\end{itemize}
To permit easy reproduction of the experiment, the authors provide  detailed instruction on how to build the active pendulum experiment. This includes:
\begin{itemize}
    \item Guidance on how to build a pendulum with solenoids for teaching/research purposes
    \item Codes and a detailed explanation for wireless data transmission between the receiver and transmitter of gyro measurements
    \item Algorithms for identifying pendulum dynamics parameters
    \item Control strategies for the solenoids for damping/pumping the swing
\end{itemize}
 The current design can be used to illustrate interesting facts related to the pendulum dynamics. For instance, the payload of the pendulum could be changed to illustrate that the frequency of oscillation does not depend on the payload mass, and the change in period of oscillation as a function of initial displacement can be used to illustrate the nonlinear characteristics of the system. The codes and algorithm in a ready to implement form are available via \href{https://github.com/AdrianStein93/Education_paper}{Github}. The authors have also created an instruction video posted on \href{https://www.youtube.com/watch?v=jvpJSOegNUs}{YouTube} to provide a step-by-step description of the assembly process for the construction of the system.\\
\\
\textbf{Github:}\\
\url{https://github.com/AdrianStein93/Education_paper}\\
\textbf{YouTube:}\\
\url{https://www.youtube.com/watch?v=jvpJSOegNUs}

\section{Acknowledgment}
The authors would like to acknowledge the support of Dr. Jesse Callanan, Youngjin Kim and Revant Adlakha in the design of the hardware. The Authors declare that there is no conflict of interest. This research received no specific grant from any funding agency in the public, commercial, or not-for-profit sectors.

\vspace{-0.1in}
\bibliographystyle{elsarticle-num}
\bibliography{Reference} 
\noindent\rule[0.5ex]{\linewidth}{1pt}
\newpage
\appendix
\section*{Appendix}
\section{Parts list}
\label{sec:05}
The following parts and tools are needed for the assembly:
\begin{itemize}
    \item Solder equipment with solder wire
    \item $3$D printer
    \item M5 x 0.8 thread cutter (or 10/32 inch)
    \item Metal saw
    \item Wood saw
    \item Wood drill
    \item Vernier caliper
    \item Drilling machine with cross screwdriver bit and wood drill and steel drill attachments
    \item Bending machine
    \item M5 and M8 wrench
    \item Pliers for the solenoid nuts
    \item Cross screwdriver
    \item 22 gauge wire
    \item Sticky tape
    \item Thin sponge material (about $20mm$ thickness)
    \item Super glue
    \end{itemize}
Table~\ref{table:mechanic_electric_misc} lists all the materials which need to be ordered.
\begin{table}
\caption{Purchased mechanical, electrical and miscellaneous material with the price from 12/13/2021.}
\centering 
\small
\begin{tabular}{l c c c c c c} 
\toprule 
 & Part & Quantity & Price ($\$$) & Vendor\\
 \midrule
 \multirow{6}{*}{Mechanic} & Ball Bearing (\(\diameter\) $8 mm$) & $2$ & 9.64 &  \href{https://www.mcmaster.com/5972K91/}{McMasterCarr} \\
 & Ball Bearing (LM5UU) (\(\diameter\) $5 mm$) & 1 & 11.99 & \href{https://www.amazon.com/uxcell-LM5UU-Linear-Bearings-Length/dp/B07H94YKQQ/ref=sr_1_4?keywords=LM5UU&qid=1639426601&sr=8-4}{Amazon}\\
 & Metal rod (\(\diameter\) $8 mm$ and $1$' length) & $1$ & $4.38$ & \href{https://www.mcmaster.com/8920K26-8920K261/}{McMasterCarr}\\
 & Metal rod (\(\diameter\) $5 mm$ and $3$' length) & $1$ & $5.63$ & \href{https://www.mcmaster.com/8920K18-8920K183/}{McMasterCarr}\\
 & Copper tube (\(\diameter\) $5/16$" and $3$' length) & $1$ & $10.95$ & \href{https://www.mcmaster.com/8967K592/}{Mc Master Carr}\\
 & Solenoid & $2$ & $113.28$ & \href{https://www.digikey.com/en/products/detail/pontiac-coil-inc/F0494A/668320}{Digi-Key}\\
 \midrule
 \multirow{6}{*}{Electric} & Transceiver set & $1$ & $19.99$ & \href{https://www.amazon.com/WayinTop-Transceiver-Regulator-ATmega328P-Communication/dp/B07ZCKMNRG/ref=sr_1_3?dchild=1\&keywords=nrf24l01+transmitter+and+receiver\&qid=1632255621&sr=8-3}{Amazon}\\
 & Gyroscope & $1$ & $5.99$ & \href{https://www.amazon.com/HiLetgo-MPU-6050-Accelerometer-Gyroscope-Converter/dp/B00LP25V1A/ref=sr_1_1_sspa?crid=253XD1LH0I730&dchild=1&keywords=mpu6050&qid=1632255725&sprefix=mou6050\%2Caps\%2C151&sr=8-1-spons&psc=1&spLa=ZW5jcnlwdGVkUXVhbGlmaWVyPUExVlhER0tMS0tZNDE2JmVuY3J5cHRlZElkPUEwNTM3NDA1MUJYWjlBWU9OREY5RCZlbmNyeXB0ZWRBZElkPUEwNzcxMjIxMU9JSTA4S001Wk40MiZ3aWRnZXROYW1lPXNwX2F0ZiZhY3Rpb249Y2xpY2tSZWRpcmVjdCZkb05vdExvZ0NsaWNrPXRydWU=}{Amazon}\\
 & Powerbank & $1$ & $19.99$ & \href{https://www.amazon.com/INIU-High-Speed-Flashlight-Powerbank-Compatible/dp/B07CZDXDG8/ref=sr_1_3?dchild=1&keywords=very+good+electronics+power+bank+5v&qid=1632331624&sr=8-3}{Amazon}\\
 & Relay Module set & $1$ & $6.19$ & \href{https://www.amazon.com/HiLetgo-Channel-optocoupler-Support-Trigger/dp/B00LW15A4W/ref=sr_1_8?dchild=1&keywords=relay+module+\%28dc30v+10a+\%29+arduino&qid=1633118701&sr=8-8}{Amazon}\\
  & $5$ A Fuse & $1$ & $4.99$ & \href{https://www.amazon.com/Standard-Blade-Fuse-Automotive-Truck/dp/B08HDHBFJB/ref=sr_1_10?keywords=5A+Fuse&qid=1638287810&qsid=136-5128116-1630355&sr=8-10&sres=B07L89K6VD\%2CB07L8JDG7R\%2CB09BT769KY\%2CB07ZFJ7M9G\%2CB07X313Q4T\%2CB08HDHBFJB\%2CB081GD1V1K\%2CB00LCIKOAU\%2CB098NCDFW1\%2CB09D9C71T7\%2CB07Q571LB3\%2CB07V2J5B6T\%2CB08L3JK17Z\%2CB09F9FYPNF\%2CB07552SD84\%2CB09HBYY1LV\%2CB07V5MYBL1\%2CB07V9WS6GM\%2CB08FYFLMPX\%2CB01NAQH2TB&srpt=FUSE}{Amazon}\\
 & Power Supply ($30$ V; $10$ A) & $1$ & $89.98$ & \href{https://www.amazon.com/4-Digital-Precision-Adjustable-Regulated-Switching/dp/B07JC6XMZ2/ref=sr_1_3?dchild=1&keywords=30v+10a+dc&qid=1618511144&sr=8-3}{Amazon}\\
 \midrule
\multirow{18}{*}{Miscellaneous} & M8x1.25 screw ($80$ mm length) & $1$ & $11.82$ & \href{https://www.mcmaster.com/91292A212/}{McMasterCarr}\\
& M$8$x$1.25$ screw ($30$ mm length) & $1$ & $12.22$ & \href{https://www.mcmaster.com/91292a149/}{McMasterCarr}\\
& M$5$x$0.8$ screw ($110$ mm length) & $1$ & $11.03$ & \href{https://www.mcmaster.com/91292A322/}{McMasterCarr}\\
& M$2.5$x$0.45$ screw ($20$ mm length) & $1$ & $6.43$ & \href{https://www.mcmaster.com/92010A026/}{McMasterCarr}\\
& M$8$ washer & $1$ & $7.14$ & \href{https://www.mcmaster.com/98689A116/}{McMasterCarr}\\
& M$5$ washer & $1$ & $3.14$ & \href{https://www.mcmaster.com/98689A114/}{McMasterCarr}\\
& M$2.5$ washer & $1$ & $1.67$ & \href{https://www.mcmaster.com/93475A196/}{McMasterCarr}\\
& M$8$x$1.25$ nut & $1$ & $6.14$ & \href{https://www.mcmaster.com/90592A022/}{McMasterCarr}\\
& M$5$x$0.8$ nut & $1$ & $1.76$ & \href{https://www.mcmaster.com/90592A095/}{McMasterCarr}\\
& M$2.5$x$0.45$ nut & $1$ & $1.94$ & \href{https://www.mcmaster.com/90592A080/}{McMasterCarr}\\
& Wood ($2$" x $6$" x $10$') & $1$ & $\approx 10$ & \href{https://www.homedepot.com/p/WeatherShield-2-in-x-6-in-x-10-ft-2-Prime-Ground-Contact-Pressure-Treated-Lumber-253921/206967800}{Home Depot}\\
& Wood ($2$" x $4$" x $8$') & $1$ & $\approx 4$ & \href{https://www.homedepot.com/p/2-in-x-4-in-x-96-in-Prime-Whitewood-Stud-058449/312528776}{Home Depot}\\
& Corner Braces ($2$" x $1$-$1$/$2$" x $2$-$3$/$4$") & $4$ & $3.36$ & \href{https://www.homedepot.com/p/Simpson-Strong-Tie-2-in-x-1-1-2-in-x-2-3-4-in-Galvanized-Angle-A23/100374944}{Home Depot}\\
& Screws \#$9$ x $1$"  & $1$ & $6.84$ & \href{https://www.mcmaster.com/90031A223/}{McMasterCarr}\\
& Screws \#$9$ x $3$"  & $1$ & $8.56$ & \href{https://www.mcmaster.com/90031A233/}{McMasterCarr}\\
& Wood ($20$" x $20$" x $1$") & $1$ & $50.59$ & \href{https://www.amazon.com/Thick-Plywood-Squares-20-Inch/dp/B07D86XLV3/ref=sr_1_3?keywords=plywood\%2B1\%2Binch\%2Bthick\%2B20&qid=1638201280&qsid=136-5128116-1630355&sr=8-3&sres=B07D85J6JS\%2CB07D86XLV3\%2CB08TLZBMP9\%2CB01MSWYF8N\%2CB088ZVL6YY\%2CB07T25FG2W\%2CB07MY8HC8L\%2CB07JB3BSYW\%2CB08T1J6PR9\%2CB07Q8DTLYS\%2CB08RJ2LRWQ\%2CB07GHX2M35\%2CB08L62VSH1\%2CB07GTQDTLR\%2CB08429ZRB1\%2CB0736P3SS7\%2CB075CTXXVS\%2CB07YTN165Y\%2CB08YJXHCZX\%2CB0971L3KX5&th=1}{Amazon}\\
& Steel sheet ($12$" x $12$" x $0.035$") & $1$ & $23.25$ & \href{https://www.amazon.com/RMP-Ga-Stainless-Steel-Sheet/dp/B00ICS72HQ/ref=sr_1_3?keywords=20+gauge+steel+sheet&qid=1636649630&qsid=136-5128116-1630355&sr=8-3&sres=B00ICS72HQ\%2CB08WJ1CWS2\%2CB08X3G4VNV\%2CB01CITA97E\%2CB07PQGDB93\%2CB00ICS74XS\%2CB00ICS715Y\%2CB0010ZXE9I\%2CB08NVB9PVS\%2CB07P958JVN\%2CB000H9RPL6\%2CB000H5W3PI\%2CB000W6ZMQY\%2CB00MONH3R4\%2CB08NV9XV5Y\%2CB00P811GZO}{Amazon}\\
& \#$9$ O-ring & $1$ & $2.92$ & \href{https://www.homedepot.com/p/DANCO-9-O-Ring-10-Pack-96726/100299667}{Home depot}\\
& Filament & $1$ & $20.99$ & \href{https://www.amazon.com/CCTREE-Printer-Filament-Accuracy-Creality/dp/B06XR7621X/ref=sr_1_4?crid=35NK4H87MNN9T&dchild=1&keywords=ender\%2B3\%2Bv2\%2Bfilament&qid=1620411075&sprefix=ender\%2B3v\%2Bfil\%\%2C189&sr=8-4&th=1}{Amazon}\\
 \bottomrule
 \label{table:mechanic_electric_misc}
 \end{tabular}
 \end{table}
The total price of the materials is: $\$$ $496.8$  (before tax). Table~\ref{table:3D_printed} shows all the $3$D printed parts needed.
\begin{table}
\caption{$3$D printed parts for the experiment.}
\centering 
\small
\begin{tabular}{l c c c c c c} 
\toprule 
Part number & Part name & Quantity\\
\midrule
Bearing bracket & \circled{$1$} & $2$\\
Housing solenoid  & \circled{$2$} & $1$\\
Adapter spacer & \circled{$3$} & $4$\\
Mid base & \circled{$4$} & $1$\\
Spacer Bearing & \circled{$5$} & $1$\\
Lower base & \circled{$6$} & $1$\\
Spacer solenoid  & \circled{$7$} & $2$\\
Payload chassis cover & \circled{$8$} & $1$\\
Payload chassis & \circled{$9$} & $1$\\
Copper tube spacer & \circled{$10$} & $2$\\
\midrule
\bottomrule
\label{table:3D_printed}
\end{tabular}
\end{table}

\begin{figure}
	\centering
	\includegraphics[width=0.95\textwidth]{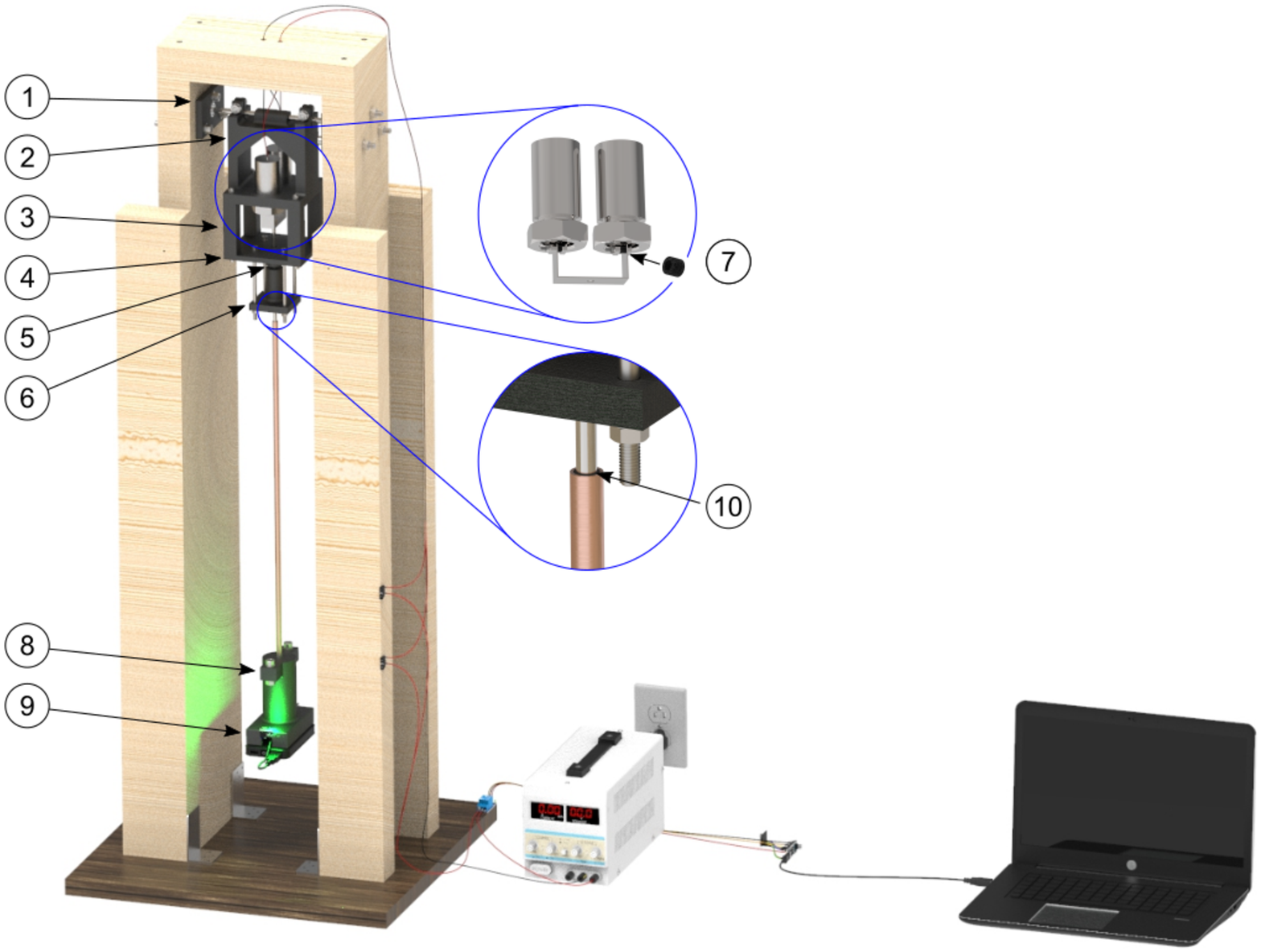}
	\caption{General setup of the pendulum with highlighting all the $3$D printed parts}
	\label{fig:04}
\end{figure}

 \href{https://www.amazon.com/Integrated-Structure-Motherboard-Carborundum-8-66x8-66x9-84in/dp/B07FFTHMMN/ref=sr_1_3?keywords=3d+printer+ender+3+v2&qid=1636078230&qsid=138-4526468-3602503&sr=8-3&sres=B07FFTHMMN\%2CB08BL41ZMY\%2CB08FSGXK46\%2CB097QR458S\%2CB08J713D9Q\%2CB09723JYZC\%2CB088GTBQWD\%2CB088BHXVLV\%2CB0932B6FS9\%2CB091TL6V94\%2CB08QZ5F1W9\%2CB08V1JMDH8\%2CB087FDTV3X\%2CB08S6LR7DQ\%2CB07H21QPTT\%2CB09B1K3RVM}{A Creality Ender 3 V2 $3$D printer} was used to print the $3$D printed parts. Figure~\ref{fig:04} illustrates the location off all $3$D printed parts for the experiment.
\section{Assembly}
\label{sec:06}
\begin{itemize}
\item The \(\diameter\) $8$ mm metal rod needs to be cut to a length of $200$ mm.
\item The \(\diameter\) $5$ metal rod needs to be cut to a length of $700$ mm. A M$5$x$0.8$ thread of $20$ mm length needs to be cut on both ends of the metal rod. (Hint: In case there is no M$5$x$0.8$ thread cutter available, a $10$/$32$ inch thread cutter can be used too).
\item Cut the steel sheet into a rectangular shape of $20$ mm x $115$ mm and drill holes at the places shown in the Figure~\ref{fig:05}. Finally the steel sheet needs to be bent into a U-shape profile (each angle is 90$^\circ$).
\begin{figure}
	\centering
	\includegraphics[width=0.90\textwidth]{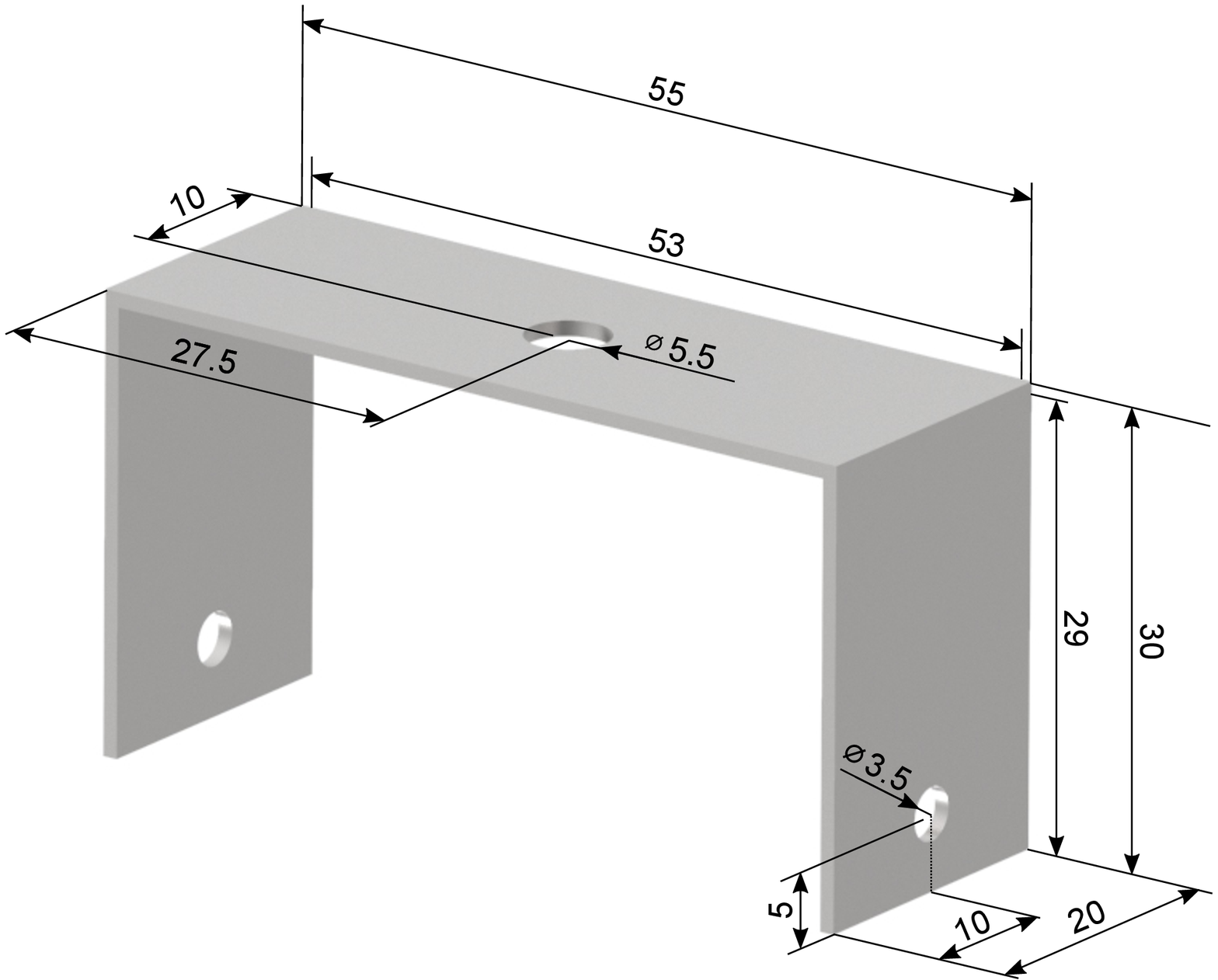}
	\caption{Steel sheet bent to a U-shape profile with bore placement instructions. All dimensions are in mm.}
	\label{fig:05}
\end{figure}
\item The copper tube needs to be cut to a length of $520$ mm.
\item The $6$” x $2$” x $10$’ wood  needs to be cut in two $1200$ mm long pieces (wood 1) and one $306$ mm long piece (wood 2). The holes illustrated in Figure~\ref{fig:06} need to be pre-drilled.
\begin{figure}
	\centering
	\includegraphics[width=0.90\textwidth]{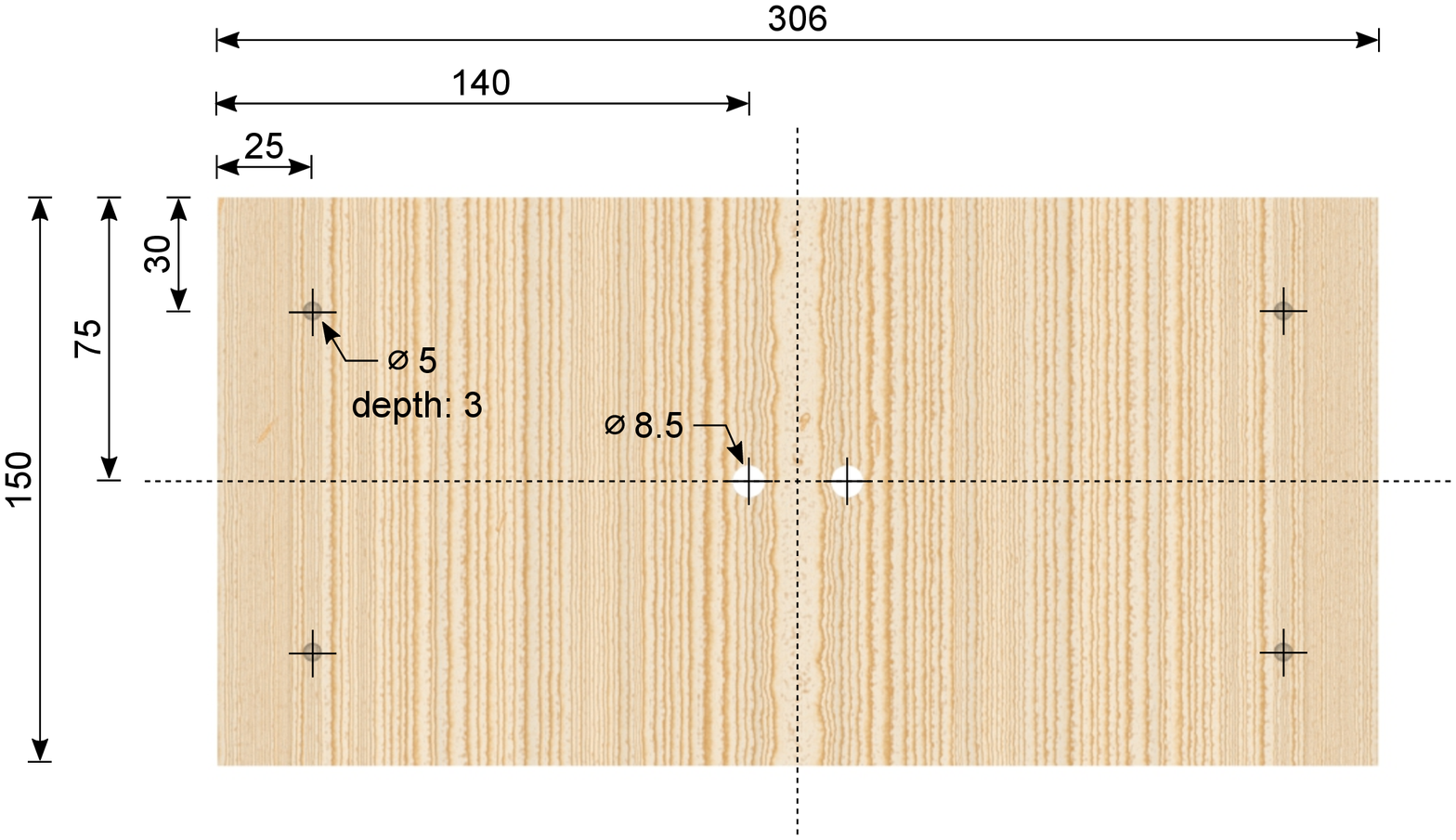}
	\caption{Drill pattern of wood $2$ (top). The depth of the \(\diameter\) $5$ mm holes is $3$ mm. All dimensions are in mm.}
	\label{fig:06}
\end{figure}
\begin{figure}
	\centering
	\includegraphics[width=0.90\textwidth]{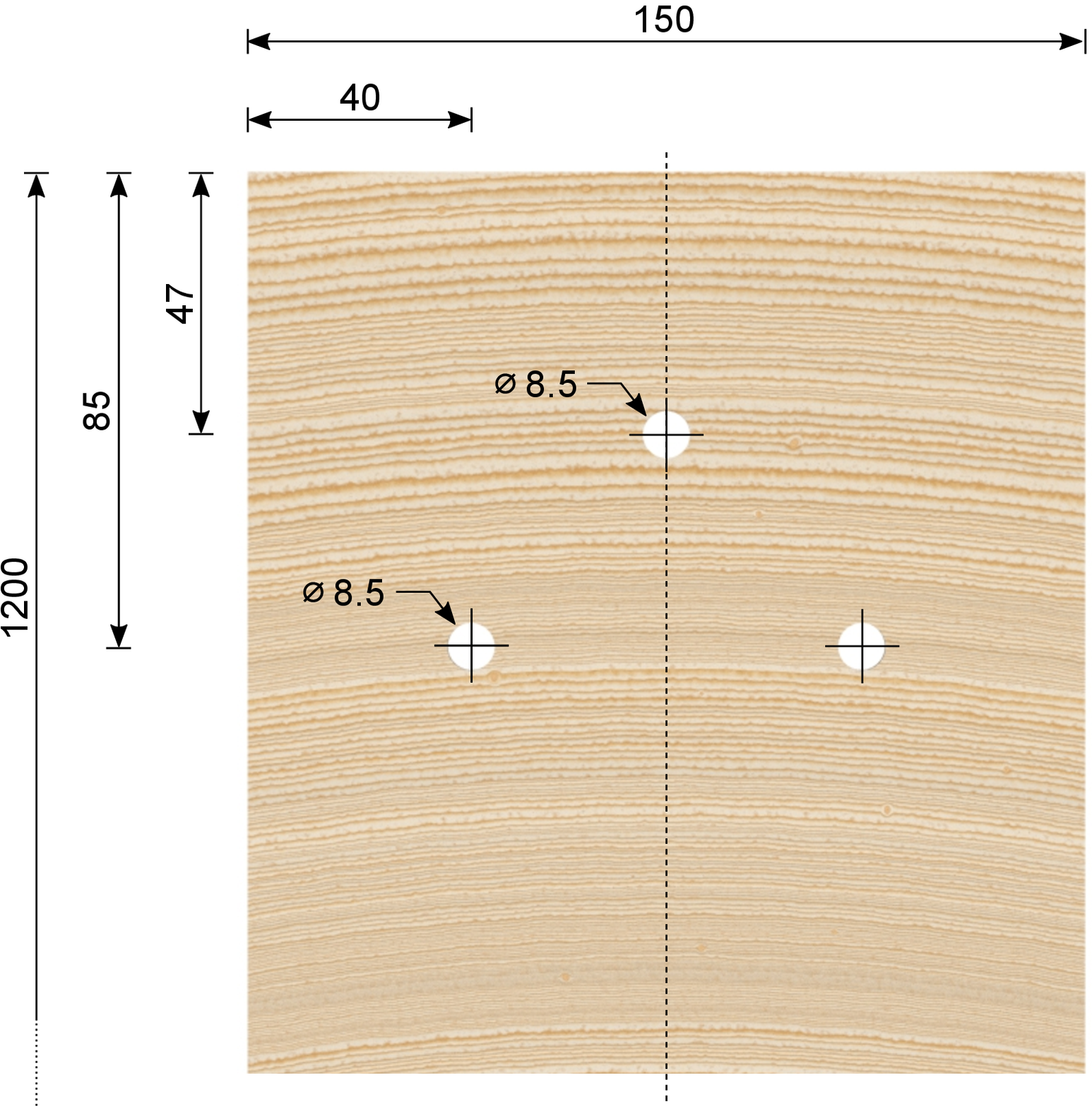}
	\caption{Drill pattern of wood $1$, where the bearing will be attached to.  All dimensions are in mm.}
	\label{fig:07}
\end{figure}
\item The wood base (wood 3) needs to be pre-drilled as follows
\begin{figure}
	\centering
	\includegraphics[width=0.90\textwidth]{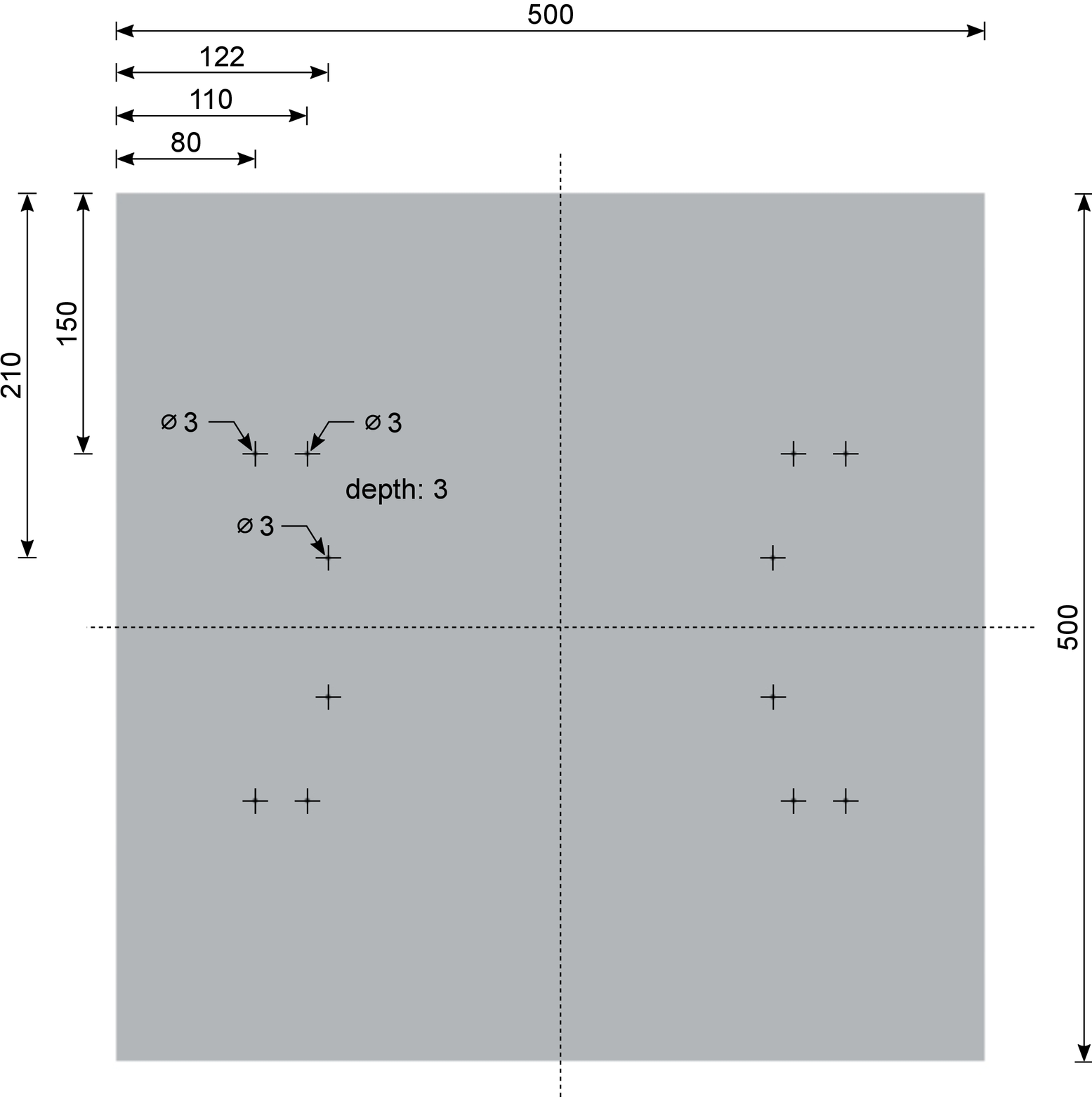}
	\caption{Drill pattern of wood $3$ (bottom). The depth of the \(\diameter\) $3$ mm holes is $3$ mm. Note: The color is grey to show the drill pattern better. All dimensions are in mm.}
	\label{fig:08}
\end{figure}
\item The two $4$" x $2$" x $8$' wood pieces need to be cut to $4$ x $1000$ mm long wood pieces. Two of them (wood $4$) and the other two (wood $5$) get pre-drilled as follows
\begin{figure}
	\centering
	\includegraphics[width=0.90\textwidth]{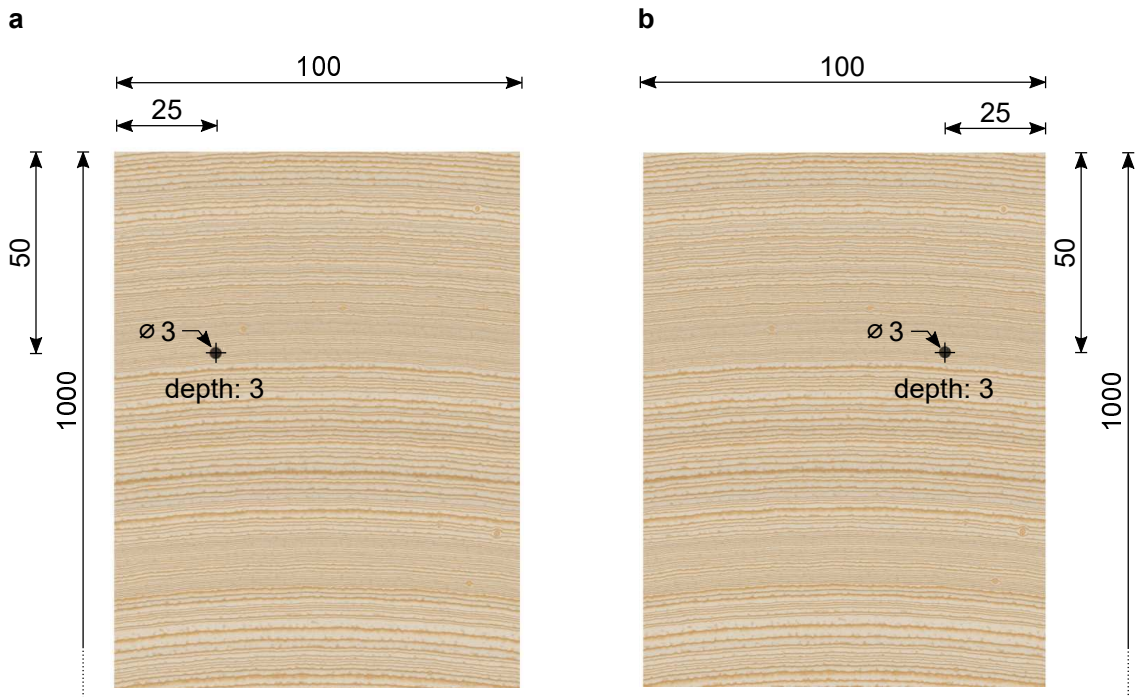}
	\caption{Drill pattern of wood 4 and 5. The depth of the \(\diameter\) $3$ mm holes is $3$ mm. a) wood 4. b) wood 5. 2 copies of each wood are needed. All dimensions are in mm.}
	\label{fig:09}
\end{figure}
\end{itemize}
The whole instruction is explained on \href{https://www.youtube.com/watch?v=jvpJSOegNUs}{YouTube}. These steps provide an instruction for the assembly:
\subsection{Step 1}
At the beginning of the assembly process, the solenoids are placed in the upper part of the $3$D printed part \circled{$2$} using $1$ $1$/$8$''-$7$ nuts. The \(\diameter\) $8$ mm metal rod gets pressed symmetric about the center through the guidance of part \circled{$2$} and is secured with $2$ x M$8$x$1.25$ screws ($30$ mm length), $4$ x M$8$ washers and $2$ x M$8$ nuts. $4$ x M$5$x$0.8$ screw ($110$ mm length) and $4$ x M$5$ washers need to be put in place in part \circled{$4$} for the next step's assembly with part \circled{$6$}. Then, part \circled{$4$} can be attached to \circled{$2$} with $4$ x \circled{$3$}, and fastened with $4$ x M$5$x$0.8$ screws ($110$ mm length), $8$ x M$5$ washers and $4$ x M$5$ nuts. 
\subsection{Step 2}
One \(\diameter\) $5$ mm linear ball bearing is pressed in the \circled{$4$}. \circled{$5$} is pressed in \circled{$4$} against the linear bearing. Simultaneously another \(\diameter\) $5$ mm linear ball bearing is pressed in \circled{$6$}. \circled{$6$} gets now pressed against \circled{$4$}, where \circled{$5$} and the $4$ pre-assembled M$5$x$0.8$ screws ($110$ mm length) function as guidance. $4$ x M$5$ washers and $4$ x M$5$ nuts are used to fasten the setup.
\subsection{Step 3}
\begin{figure}
	\centering
	\includegraphics[width=0.90\textwidth]{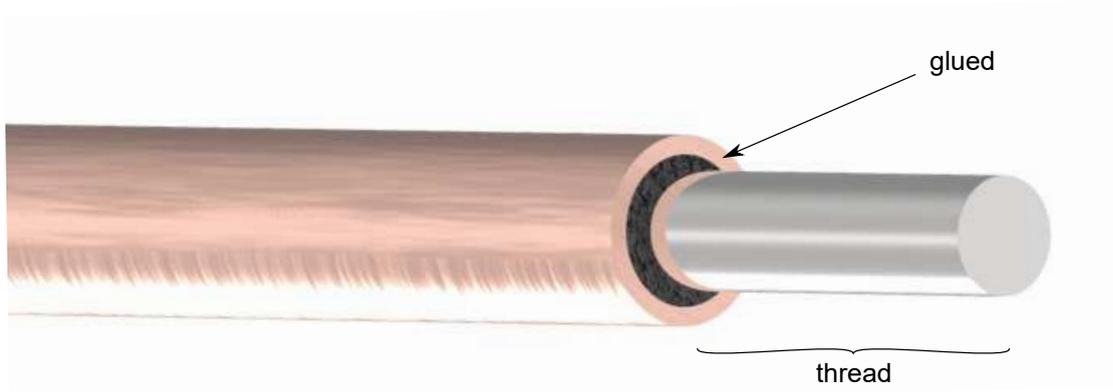}
	\caption{Copper tube spacer. The spacer (black) is $3$D printed.}
	\label{fig:10}
\end{figure}

\begin{figure}
	\centering
	\includegraphics[width=0.90\textwidth]{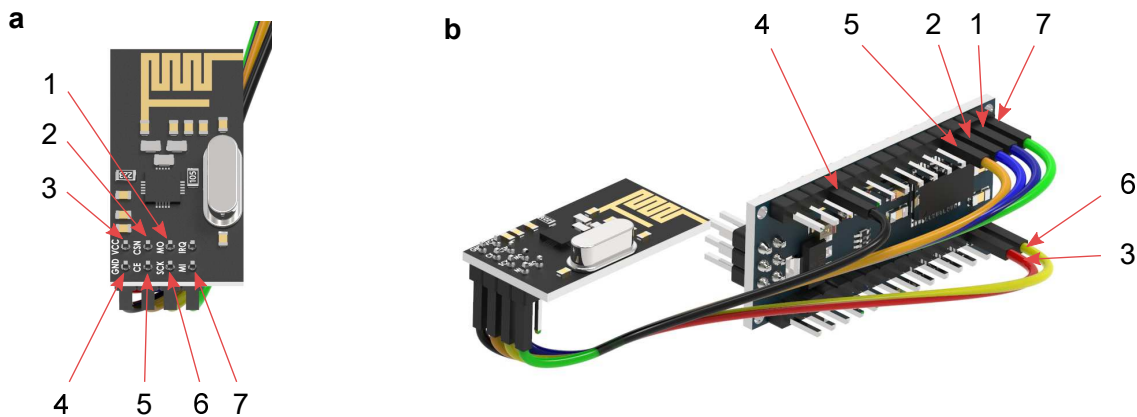}
	\caption{Wiring Transmitter and Arduino Nano. \textbf{Transmitter}:  $1$ - MOSI, $2$ - CSN, $3$ - VCC, $4$ - GND, $5$ - CE, $6$ - SCK, $7$ - MISO; \textbf{Arduino Nano}: $1$ - D$11$, $2$ - D$10$, $3$ - $3$V$3$, $4$ - GND, $5$ - D$9$, $6$ - D$13$, $7$ - D$12$.}
	\label{fig:11}
\end{figure}

\begin{figure}
	\centering
	\includegraphics[width=0.90\textwidth]{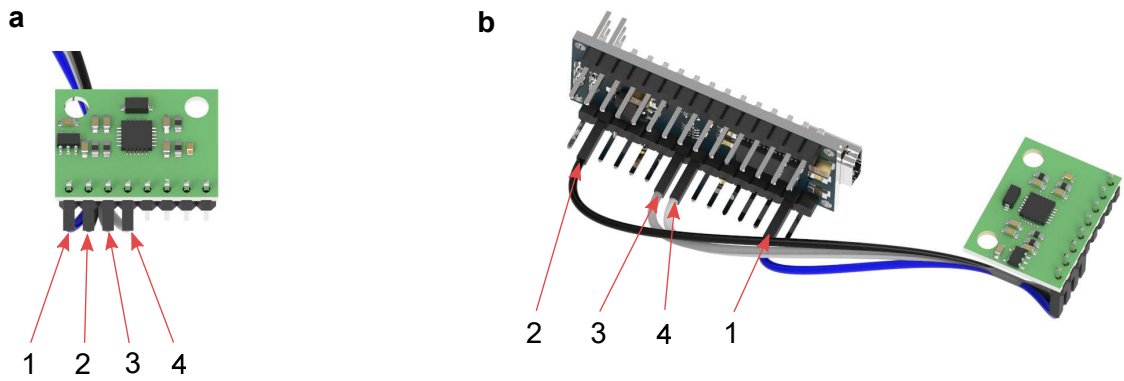}
	\caption{Wiring Gyroscope and Arduino Nano. \textbf{Gyroscope}:  $1$ - VCC, $2$ - GND, $3$ - SCL, $4$ - SDA; \textbf{Arduino Nano}: $1$ - $3$V$3$, $2$ - GND, $3$ - A$5$, $4$ - A$4$.}
	\label{fig:12}
\end{figure}

$1$ x \circled{$10$} gets slid over the \(\diameter\) $5$ mm metal rod until the end of the thread from either side. The copper tube is slid over \circled{$10$} until the ends of \circled{$10$} and the copper tube match (as illustrated in Figure~\ref{fig:10}). Super glue is used to create a stiff connection between the metal rod, \circled{$10$} and the copper tube as shown in Figure~\ref{fig:10}. From the other end of the rod, another \circled{$10$} is slid over the metal rod and guided into the copper tube until it is completely in between the metal rod and the copper tube. Super glue is used again to create a stiff connection.
\circled{$8$} is attached to the \(\diameter\) $5$ mm metal rod  by $2$ x M$5$x$0.8$ nuts and $2$ x M$5$ washers from each end of \circled{$8$} respectively. \circled{$8$} is connected to \circled{$9$} by $2$ x M$8$x$1.25$ screws ($30$ mm length), $4$ x M$8$ washers and $2$ x M$8$ nuts. \circled{$9$} houses an Arduino Nano, and holds a MPU$6050$ gyroscope and a transmitter. A powerbank supplies the components with power. The gyroscope and the transmitter are attached to the $3$D printed body with sticky tape and a spongy material is used between the sensors and \circled{$9$}. A mini USB cable connects the Arduino Nano to the powerbank. The wiring between the Arduino Nano, gyroscope and the transmitter can be found in Figure~\ref{fig:11} and Figure~\ref{fig:12}.
Note: Cable $3$ of Figure~\ref{fig:11} and cable $1$ of Figure~\ref{fig:12} need to be connected by soldering and then plugged to the Arduino Nano. Figure~\ref{fig:11} shows how the final sensor part looks.
\begin{figure}
	\centering
	\includegraphics[width=0.95\textwidth]{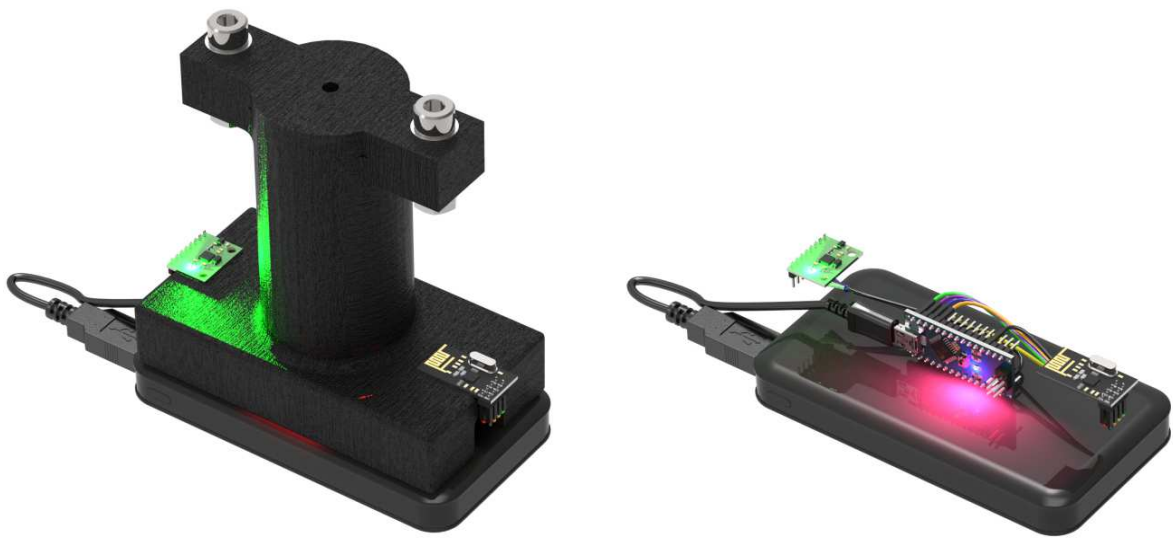}
	\caption{Assembled payload chassis cover and payload chassis with transmitter, Arduino Nano and Gyroscope}
	\label{fig:13}
\end{figure}
\subsection{Step 4}
$1$ x \#$9$ O-ring needs to be placed on each solenoid pin. The U-shaped profile gets plugged into the solenoids moving pins. $2$ x part \circled{$7$} need to be placed between the U-profile and the solenoid pins. $2$ x M$2.5$x$0.45$ ($20$ mm length) screws, $4$ x M$2.5$ washers and $4$ x M$2.5$ nuts are used to fasten the U-shaped profile to the solenoid pins. The \(\diameter\) $5$ mm metal rod is slid through \circled{$6$}, \circled{$4$} and both linear bearings. On the metal rod thread a M5x0.8 nut and a M5 washer need to be placed on the rod to hold it back, before the rod slides through the U-shaped profile and gets fastened by another M$5$x$0.8$ nut and a M$5$ washer. A spongy material needs to be placed between \circled{$9$} and the U-shaped profile to dampen the impact of the solenoid's movement.
\subsection{Step 5}
Both wood $2$ are being put on the wood $3$ so that the assembled pendulum is placed as centered. $3$ x M$8$x$1.25$ screws ($80$ mm length), $6$ x M$8$ washers and $3$ x M$8$ nuts are used to attach \circled{$1$} to each wood $2$. $2$ x \(\diameter\) $8$ mm Ball Bearings are pressed into \circled{$1$}. Then both wood $2$ can be put together with the assembled pendulum from both sides, where the bearing function as a guidance. Wood $1$ needs to be put on the top and fastened with $4$ x \#$9$x$3$ screws in total using the pre-drilled holes as guidance. $2$ x wood $4$ and $2$ x wood $5$ get put against the assembled structure, so that the inner parts align. $4$ x \#$9$x$3$ screws are used in total to fasten wood $4$ and wood $5$ to the assembled structure by using the pre-drilled holes as guidance. $4$ x corner braces with each $8$ x \#$9$x$1$ screws are used to stabilize wood $4$ and wood $5$ by being placed on the inner side. Finally, $12$ x \#$9$x$3$ screws are screwed from the bottom side of wood $3$ by using the pre-drilled whole as orientation points to stabilize the whole structure.
 \subsection{Step 6}
\begin{figure}
	\centering
	\includegraphics[width=0.90\textwidth]{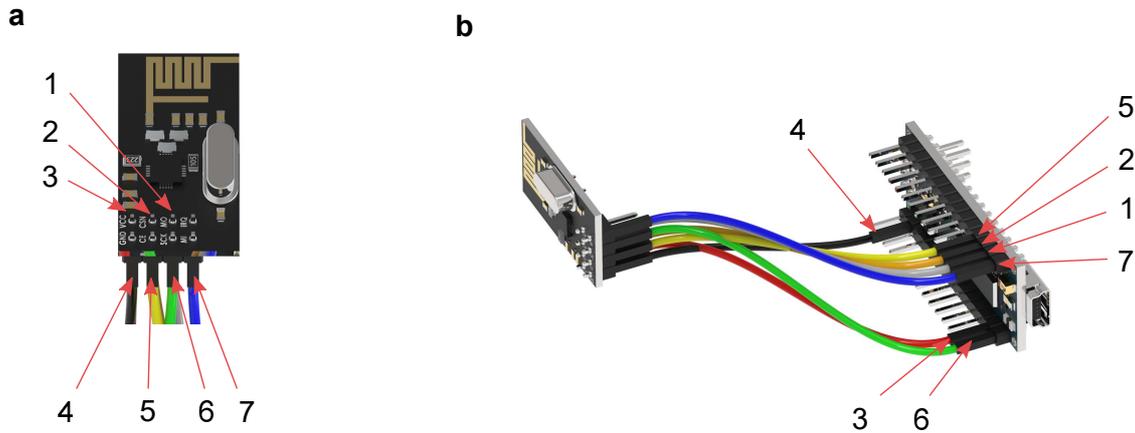}
	\caption{Wiring Receiver and Arduino Nano. \textbf{Receiver}:  $1$ - MOSI, $2$ - CSN, $3$ - VCC, $4$ - GND, $5$ - CE, $6$ - SCK, $7$ - MISO; \textbf{Arduino Nano}: $1$ - D$11$, $2$ - D$10$, $3$ - $3$V$3$, $4$ - GND, $5$ - D$9$, $6$ - D$13$, $7$ - D$12$.}
	\label{fig:14}
\end{figure}

\begin{figure}
	\centering
	\includegraphics[width=0.90\textwidth]{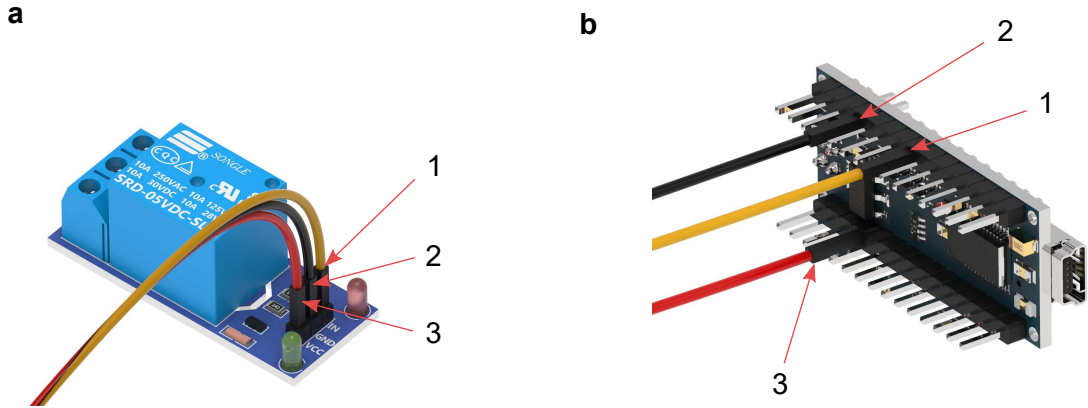}
	\caption{Wiring Relay and Arduino Nano. \textbf{Relay module}: $1$ - Signal input, $2$ - DC-, $3$ - DC+; \textbf{Arduino Nano}: $1$ - D$5$, $2$ - GND, $3$ - $5$ V.}
	\label{fig:15}
\end{figure}

 Connect the power cord of the Power Supply to a socket and place the relay module on wood $3$. Insert the mini USB cable in the computer and connect it to the Arduino Nano which is used to receive the data. Wire the receiving Arduino Nano to the receiver module as prescribed in Figure~\ref{fig:14}
Use $2$ x $5$ A fuses and place them on wood $4$. Fix them with sticky tape. Wire the receiving Arduino Nano to the relay module as prescribed in Figure~\ref{fig:15}.
Solder $22$ gauge wire to each black wire coming out of the solenoid to extend them. Use the black crocodile clip which came with the power supply and grab both extended black cables from the solenoids and plug the other end in the power supply. Connect the red crocodile clip to the power supply and connect it to a $22$ gauge wire. Connect this wire to the ``COM'' output of the relay. Take two $22$ gauge wires and insert both of them into the ``NO'' (Normally open) output of the relay module. The other ends of both wires need to be soldered on the each fuse respectively. Take again $22$ gauge wires and solder each of them on the remaining spot of the fuses. The other end of the wires need to be soldered to the red cables of solenoid.
\section{Data acquisition}
\label{sec:data_acquisition}
We decided to use the Arduino IDE software to control the Arduino Nanos. Two codes need to be written for the transmitter and receiver respectively which can be downloaded with the following \href{https://github.com/AdrianStein93/Education_paper}{Github link}. Download the ``MPU6050\_light.h'' library which is used to get the data from the gyroscope and then download the ``nRF24L01.h'' library which is used to send and receive data with the nRF24L01 modules (transmitter \& receiver) and embed them in the Arduino IDE environment. For the experiment there are several requirements which need to be taken into account
\begin{itemize}
\item 1. Wireless transmission of the angular velocity $\dot{\Phi}$ with a reasonable sampling frequency
\item 2. Robustness to the magnetic field caused by the solenoids
\item 3. Robustness to the hard impacts caused by the solenoids
\item 4. Remote calibration of the gyroscope
\end{itemize}
1. The difficulty is that 4. is causing a trade-off with 1. If the transmitter can be calibrated at any time instant, then the sampling frequency drops because it needs to be enabled to ``listen'' during every void loop. We will describe our algorithm on how we achieved a reasonable sampling frequency.\\
\\
\noindent
2. Instead of a nRF24L01 receiver module with an antenna, a nRF24L01 receiver without an antenna is used. During conducted experiments the authors found out that an antenna actually gets influenced by the magnetic field and was harming the wireless transmission.\\
\\
\noindent
3. A low-pass filter for the angular velocity is implemented because the impacts caused vibrations and influence the angular velocity. A detailed explanation is provided in subsection~\nameref{subsec:low_pass_filter}.\\
\\
\noindent
4. After the transmitter code is uploaded the transmitter is just ``listening''. The time (in ms) and the angular velocity (in deg/s) are sent with a frequency of $\approx 5$ Hz and the low sampling frequency shows that the receiver needs to be calibrated. After a successful calibration the sampling frequency is $\approx 116$ Hz. After $100$ s have passed the the sampling frequency drops again to $\approx 5$ Hz, which shows the user that another calibration is needed and the transmitter is enabled to ``listen'' again in order to receive the calibration command.\\
\\
\noindent
Note: The transmitter code needs to uploaded just once to the Arduino Nano by using the computer. Based on the received data, the receiving Arduino decides if the solenoids are activated or deactivated.
\subsection{Predicting angular velocity and determining angle}
\label{subsec:Phi_dot_and_Phi}
\begin{figure}
	\centering
	\includegraphics[width=0.95\textwidth]{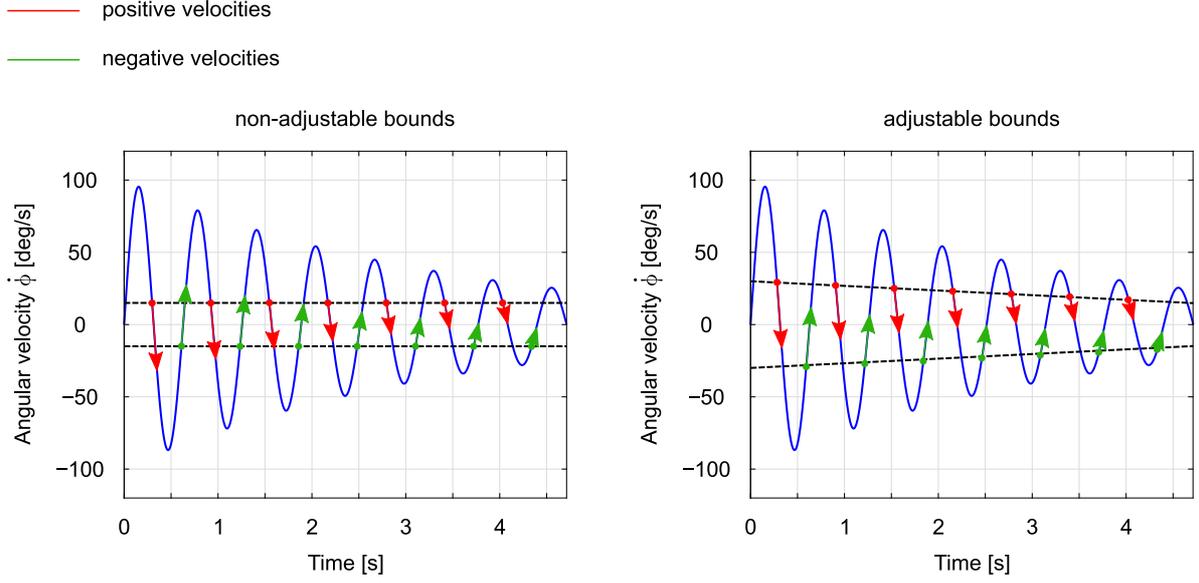}
	\caption{Difference between non-adjustable (left figure) and adjustable (right figure) bounds to capture $\dot{\Phi} = 0$ $^\circ$/s. Over time the bounds are shrinking as illustrated in the right figure. This is an example of adjustable bounds for an attenuation experiment.}
	\label{fig:16}
\end{figure}

\begin{figure}
	\centering
	\includegraphics[width=0.90\textwidth]{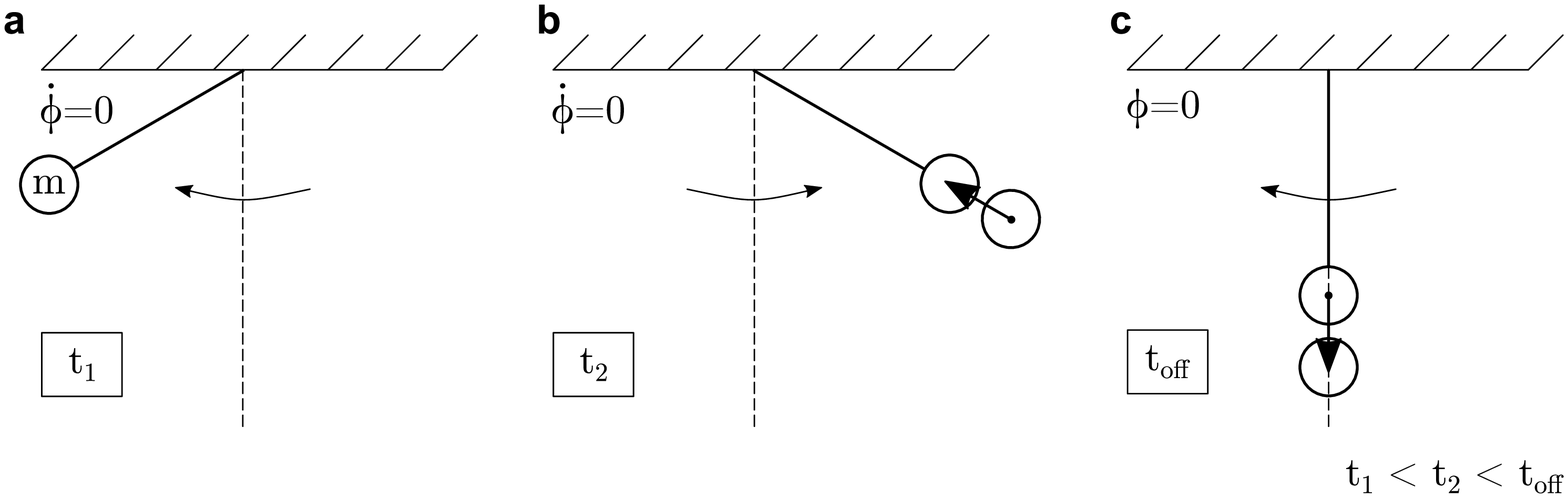}
	\caption{Strategy to capture the angular displacement $\Phi=0$ $^\circ$ by purely measuring the angular velocity $\dot{\Phi}$. a) $\dot{\Phi}_{n+1}$ got captured at $t_1$. b) $\dot{\Phi}_{n+1}$ got captured at $t_2$. c) $1$/$4$ of the time period has passed and the solenoid is being deactivated.}
	\label{fig:17}
\end{figure}

We need to account for the dead time which is the delay from the instant when $\dot{\Phi} = 0$ to the activation of the solenoid. We estimated for our setup that the dead time was $t_{dead,on} = 0.09$ s. Since one cannot exactly measure the time instant when $\dot{\Phi} = 0$, and to account for the dead time, we use linear extrapolation to estimate the zero crossing time for $\dot{\Phi}$ using the equation:
\begin{align}
    \dot{\Phi}_{n+1} = \frac{\dot{\Phi}_n-\dot{\Phi}_{n-1}}{t_n-t_{n-1}}t_{dead,on} + \dot{\Phi}_n
\end{align}
where $\dot{\Phi}_{n+1}$ is the predicted angular velocity, $\dot{\Phi}_n$ and $\dot{\Phi}_{n-1}$ are the angular velocities at times $t_n$ and $t_{n-1}$ respectively. It is obvious that time instants where $\dot{\Phi}_{n+1} = 0$ $^\circ$/s are difficult to catch. Therefore a lower and upper bound needs to be applied around $ 0$ $^\circ$/s and we consider:
\begin{align}
    \dot{\Phi}_l < \dot{\Phi}_{n+1} < \dot{\Phi}_u
\end{align}
holds. $\dot{\Phi_l}$ is the lower and $\dot{\Phi_u}$ the upper bound respectively. However, since we are interested in an attenuation of the angle, one can conceive of a scenario where  $\dot{\Phi_l} < \dot{\Phi}_{n+1} < \dot{\Phi_u}$, i.e., the angular velocity lies within the bounds used to forecast the time for the zero crossing of $\Phi$. To prevent such a scenario we tighten the bounds over time as illustrated in Figure~\ref{fig:16}.
For our attenuating experiment we start with $\dot{\Phi}_l = -50$ $^\circ$/s and $\dot{\Phi}_u = 50$ $^\circ$/s. The final time is $20$ s and the bounds are $\dot{\Phi}_l = -45$ $^\circ$/s and $\dot{\Phi}_u = 45$ $^\circ$/s. For any time instant in between the start and final time we linearly interpolate the bounds. For catching $\Phi = 0^\circ$ precisely we cannot rely on the angular velocity measurement because even when using a low-pass filtered signal the impact caused by the solenoids create unpredictable peaks which can't be used for an integration over time. As mentioned in section~\nameref{sec:introduction} the natural frequency is changing when the solenoids are activated compared to a deactivated case. Furthermore, the natural frequency is a function of the magnitude of angular displacements and decreases with an increase in the magnitude of displacement. After letting the payload swing the instant where $\dot{\Phi} = 0$ $^\circ$/s, would provide half a time period of a mixed system (activated \& deactivated solenoids). However the natural frequencies are reasonably close, so half of this time period provides a satisfying estimate of the time when the solenoids need to be deactivated. Figure~\ref{fig:17} illustrates the described strategy.
The dead time to deactivate the solenoids is $t_{dead,off} = 0.1$ s, where gravity is the only active force. It should be noted that the release dynamics of the solenoid on the active pendulum setup is not only a function of gravity but the centrifugal force as well, which reduces the deactivation dead time. Assuming $t_1$ and $t_2$ are two consecutive time instants when the angular velocity is zero. The time to deactivate the solenoid is given by the equation:
\begin{align}
    t_{off} = \frac{t_2-t_1}{2} + t_2 - t_{dead,off}. \label{eq:t_Phi_off}
\end{align}
We found that if $t_{dead,off}$ in Eq.~\eqref{eq:t_Phi_off} is set to $0$ s we observed experimental results match the simulation results. This can be explained by the fact that the centrifugal force is the greatest when the solenoid is deactivated and reduces the transition time. However, for the pumping up the swing experiment, the solenoid is deactivated when the centrifugal force is zero and the impact of the gravity force progressively decreases, mandating the inclusion of the transition time of the solenoid. Note that for the next iteration $t_{1,new}=t_2$ and $t_2$ will be the new measured time instant. For the pumping the swing experiment, the lower bounds for determining $\dot{\Phi}$ were $\dot{\Phi}_l=-10$ $^\circ$ and $\dot{\Phi}_u=10$ $^\circ$ at the initial time. At the final time of $t=40$ s, the bounds are $\dot{\Phi}_l=-15$ $^\circ$ and $\dot{\Phi}_u=15$ $^\circ$. The delay for contracting the solenoids is $t_{dead,on} = 0.1$ s, so the equation is:
\begin{align}
    t_{on} = \frac{t_1+t_2}{2} + t_1 - t_{dead,on}\label{eq:t_Phi_on}
\end{align}
\section{Procedure of an experiment}
The following steps should help the reader understand how an experiment is conducted. Turn on the power module and set it to 24V.
\begin{enumerate}
    \item Connect the computer to the transmitting Arduino and upload the transmitter code. Disconnect the computer from the Arduino.
    \item Connect the computer to the receiving Arduino and upload the receiver code. Open its Serial Monitor. 
    \item Place the pendulum in a vertical down position. The user is asked to choose between ``1'' for calibration or ``2'' for starting an experimental testing. Concurrently, the sampling rate is visualized. During calibration, the pendulum should be undisturbed. If the sampling is not above $100$ Hz, you need to calibrate the device and send ``1''. The process takes about $10$ s. If the calibration is successful the user gets notified and can choose Start with the ``2'' option. If the calibration wasn't successful please try to calibrate again and end a ``1''.
    \item A successful calibration for our setup a resulted in a sampling frequency in between $110$ and $125$ Hz. You are then asked to displace the pendulum in a $-90$ $^\circ$ position and hold it.
    \item  A countdown will run up to $100\%$ and a ``Let go!'' message will be displayed. This is the moment you need to release the pendulum.
\end{enumerate}
\end{document}